\documentclass[a4paper,fleqn]{cas-dc}

\usepackage[numbers]{natbib}
\usepackage{multirow}
\usepackage{makecell}
\usepackage{float}
\usepackage[ruled,vlined,linesnumbered]{algorithm2e}

\newcommand{\revise}[1]{\textcolor{black}{#1}}
\newcommand{\newrevise}[1]{\textcolor{black}{#1}}
\def\tsc#1{\csdef{#1}{\textsc{\lowercase{#1}}\xspace}}
\tsc{WGM}
\tsc{QE}
\tsc{EP}
\tsc{PMS}
\tsc{BEC}
\tsc{DE}
\begin{document}
\let\WriteBookmarks\relax
\def\floatpagepagefraction{1}
\def\textpagefraction{.001}
\shorttitle{Hybrid Quantum Annealing for Arctic Ship Routing Optimization}
\shortauthors{T. Kit et~al.}

\title [mode = title]{Hybrid Quantum Annealing Approach for High-Dimensional and Multi-Criteria Constrained Quadratic Optimization in Arctic Ship Routing}                      
\tnotemark[1,2]

\tnotetext[1]{This research was supported by ‘Quantum Information Science R\&D Ecosystem Creation’ through the National Research Foundation of Korea(NRF) funded by the Korean government (Ministry of Science and ICT(MSIT))(No. 2020M3H3A1110365).}

\tnotetext[2]{This research was supported by D-Wave leap quantum cloud service was supported by the Chungbuk Quantum Research Center at Chungbuk National University. A. R. and K. K., were supported by the MSIT(Ministry of Science and ICT), Korea, under the ITRC(Information Technology Research Center) support program(IITP-RS-2024-00437284) supervised by the IITP(Institute for Information \& Communications Technology Planning \& Evaluation. A. R. and K. K. were supported by Creation of the quantum information science R\&D ecosystem(based on human resources) (Agreement Number) through the National Research Foundation of Korea(NRF) funded by the Korean government (Ministry of Science and ICT(MSIT)) (RS-2023-00256050).}

\author[1]{Tara Kit}[type=author,
                        auid=000,bioid=1,
                        prefix=,
                        role=,
                        orcid=0009-0008-5240-015X]

\credit{Conceptualization, Methodology, Validation, Visualization, Software, Writing - review \& editing}

\affiliation[1]{organization={Department of AI Convergence, Pukyong National University},
                addressline={Nam-gu}, 
                city={Busan},
                postcode={48513}, 
                country={South Korea}}
                
\affiliation[2]{organization={Chungbuk Quantum Research Center, Chungbuk National University},
                addressline={Chungcheongbuk-do}, 
                % city={},
%               citysep={}, % Uncomment if no comma needed between city and postcode
                postcode={28535–28644}, 
                state={},
                country={South Korea}}
\author[1]{Kimsay Pov}[orcid=0009-0006-8391-3869]
\credit{Software, Writing - review \& Validation}
\author[1]{Myeongseong Go}[orcid=0009-0002-7292-9786]
\credit{Review \& Validation}
\author[1]{Leanghok Hour}[orcid=0009-0001-5670-7481]
\credit{Review \& Validation}
\author[3]{Arim Ryou}[orcid=0009-0008-8181-5055]
\credit{Review \& Validation}
\author[2,3]{Kiwoong Kim}[orcid=0000-0003-1195-5681]
\credit{Review \& Validation}
\author[4]{Tae-Kyung Kim}[orcid=0000-0001-9962-1066]
\credit{Review \& Validation}
\author[1]{Youngsun Han}[orcid=0000-0001-7712-2514]
\cormark[1]
\ead{youngsun@pknu.ac.kr}

\credit{Supervision, Validation, Visualization, Writing - review \& editing}

\affiliation[3]{organization={Department of Physics, Chungbuk National University},
                addressline={Chungcheongbuk-do}, 
                postcode={28535}, 
                postcodesep={}, 
                city={},
                country={South Korea}}

\affiliation[4]{organization={Department of Management Information Systems, Chungbuk National University},
                addressline={Seowon-Gu}, 
                postcode={28644}, 
                postcodesep={}, 
                city={Cheongju},
                country={South Korea}}

\cortext[cor1]{Corresponding author}
% \cortext[cor2]{Principal corresponding author}
% \fntext[fn1]{This is the first author footnote, but is common to third
% \fntext[fn2]{Another author footnote, this is a very long footnote and
%   it should be a really long footnote. But this footnote is not yet
%   sufficiently long enough to make two lines of footnote text.}

% \nonumnote{This note has no numbers. In this work we demonstrate $a_b$
%   the formation Y\_1 of a new type of polariton on the interface
%   between a cuprous oxide slab and a polystyrene micro-sphere placed

\begin{abstract}
The opening of Arctic sea routes presents unprecedented opportunities for global trade but poses significant operational and computational challenges due to the dynamic nature of sea-ice conditions. 
This study formulates a multi-criteria Arctic route optimization problem that integrates Copernicus Marine Environment Monitoring Service (CMEMS) variables into a Constrained Quadratic Model (CQM) and solves it using D-Wave's hybrid quantum-classical solver. 
We benchmark the feasibility and scalability of this approach against classical Mixed-Integer Quadratic Programming (MIQP) solvers such as Gurobi and CPLEX. 
Results show that the CQM formulation achieves feasible solutions with stable runtimes as quadratic density increases, \revise{demonstrating 10 to 100 times faster convergence and reduced computational time compared with classical solvers, while also improving route smoothness by approximately 10\% and reducing total length by approximately 1\%}. This reflects the effectiveness of the hybrid quantum annealing approach for Arctic routing problems.

\end{abstract}

% \includegraphics{figs/cas-grabs.pdf}

% \begin{highlights}
% \item Research highlights item 1
% \item Research highlights item 2
% \item Research highlights item 3
% \end{highlights}

\begin{keywords}
Arctic shipping routes \sep Quantum annealing \sep Constrained quadratic model \sep Mixed-integer quadratic programming \sep Sea ice dynamics \sep Route planning
\end{keywords}

\maketitle

\section{Introduction}

The accelerated retreat of Arctic sea ice has reopened high-latitude passages such as the Northern Sea Route (NSR), Northwest Passage (NWP), and Transpolar Sea Route (TSR), potentially reducing voyage distance and fuel consumption by up to 40\%~\cite{Zhang2025_NatureCO2,Wang2025_ShippingPolicy,Kozera2025_Grassroots}. 
Despite this potential, current Arctic maritime traffic remains highly concentrated along a few corridors \cite{Rodriguez2025_AIS} and dominated by bulk carriers rather than container vessels \cite{Poo2024_TradeType}. 
Reports indicate that geopolitical tensions, limited port infrastructure, and unpredictable ice still discourage major carriers from routine Arctic transits~\cite{ROVENSKAYA2024103446}. 

Satellite-based studies reveal that ice thickness, concentration, and drift velocity are key determinants of navigable windows \cite{Zhang2024_OceanEng,Aksenov2023_Risks}. 
However, most existing routing frameworks employ simplified climatological averages or static ice charts that fail to represent real-time variability \cite{Theocharis2023_Frontiers,Kotovirta2023_Review}. 
These simplifications introduce uncertainty into risk and cost estimation, motivating the integration of high-resolution CMEMS data to improve route feasibility and safety modeling.

Modeling this routing task as an optimization problem introduces \revise{significant} computational complexity. 
When sea-ice variables and curvature penalties are incorporated, the formulation becomes a dense MIQP, a class of NP-hard problems that scale exponentially with quadratic constraint density~\cite{Quirante2024_MIQP,AlKhayyal2022_Piecewise,10.1007/978-3-319-93031-2_43,Quirynen2024_ControlMIQP}. 
While classical solvers such as Gurobi and CPLEX excel at structured optimization, their runtime and memory usage grow prohibitively large under dense and dynamic Arctic constraints. 
Physical ship-ice interaction studies confirm that resistance is a non-linear function of speed, floe size, and ice thickness, further validating the use of quadratic modeling terms \cite{Huang2020_ShipIceCFD}. 
Consequently, the routing task is best expressed as an MIQP, where linear terms represent environmental and distance costs, and quadratic terms capture directional smoothness and structural consistency. 
This structure provides the mathematical expressiveness necessary to model sea-ice resistance and curvature penalties that cannot be linearized without sacrificing realism.

Another fundamental design aspect is the spatial discretization of the Arctic Ocean surface. 
Conventional latitude-longitude grids suffer from polar convergence and discontinuity across the antimeridian, creating inconsistencies in neighborhood connectivity. 
To overcome these challenges, this work employs the H3 hexagonal hierarchical grid system~\cite{Uber2018H3}, which provides near-equal-area tessellation, seamless global coverage, and efficient adjacency indexing. 
Each H3 cell represents a fixed-area oceanic unit from which environmental attributes such as sea-ice thickness, age, velocity, and concentration are extracted from Copernicus CMEMS datasets. 
This enables physically consistent graph construction even near the pole, ensuring scalable discretization across resolutions. 
The H3 framework also facilitates adjacency-based cost propagation and curvature analysis, forming a natural foundation for the MIQP route optimization model.

To address the computational bottlenecks of MIQP solving, this study adopts the CQM formulation, which is executed on D-Wave’s hybrid quantum-classical solver. 
Unlike penalty-based quadratic unconstrained binary optimization \newrevise{(QUBO) methods~\cite{Glover2022QUBO}}, the CQM directly encodes equality and inequality constraints in integer space, enhancing feasibility and solver interpretability \cite{DWave2025_CQMDocs}. 
Recent research shows that hybrid solvers can efficiently handle dense quadratic objectives and achieve near-optimal feasibility across large combinatorial spaces~\cite{Pelofske2025_Hybrid,Szal2025_TransNav}.
This research~\cite{Quinton2025_QAReview} benchmarks D-Wave's quantum annealing hybrid solvers against classical optimizers like CPLEX and Gurobi across optimization classes such as binary quadratic programming (BQP) and mixed-integer linear programming (MILP), showing competitive performance in BQP for escaping local minima via quantum tunneling, with applications to NP-hard problems in routing and navigation. Employing QUBO/Ising formulations and penalty-based constraint handling, the study evaluates solution quality and time on synthetic and real-world unit commitment scenarios, underscoring quantum annealing's advantages for quadratic binary problems akin to maritime path optimization in icy waters, while noting challenges in scalability, constraint complexity, and the value of hybrid partitioning for larger instances like Arctic route planning.
Quantum annealing optimization has also demonstrated promising scalability in maritime and logistics domains, though its performance remains problem-dependent~\cite{10632778,10.1145/3760622.3760626}.

The main contributions of this work are summarized as follows. 
This research introduces a hybrid quantum annealing formulation that bridges environmental modeling with constrained optimization:

\begin{enumerate}
    \item Multi-criteria objective formulation:
    A physically grounded cost function integrating sea-ice thickness, age, drift velocity, and curvature penalties to represent both navigational efficiency and environmental risk mitigation.

    \item High-resolution H3 Arctic graph model:
    A data-enriched H3 hexagonal discretization of the Arctic Ocean using CMEMS variables, ensuring continuous, area-preserving representation across the 180° meridian.

    \item Comparative solver benchmarking:
    A quantitative evaluation showing that D-Wave’s CQM solver maintains runtime stability and feasibility as quadratic density increases, outperforming classical solvers in scalability and practical applicability.
\end{enumerate}

\revise{A distinctive contribution of this study lies in the integration of environmental realism and computational scalability within a unified hybrid optimization framework. By coupling multi-variable CMEMS sea-ice datasets (thickness, age, drift, and concentration) with a high-resolution H3 hexagonal discretization, and solving the resulting multi-criteria MIQP formulation through D-Wave’s hybrid CQM solver, the framework achieves measurable improvements in both computational and physical performance.
Compared to classical solvers such as Gurobi and CPLEX, the proposed approach attains one to two orders of magnitude faster convergence, obtaining feasible Arctic routes within 5 to 30s, while maintaining comparable or better optimality. The resulting trajectories exhibit approximately 10\% smoother curvature and 1\% shorter total distance.
These quantitative enhancements confirm that the hybrid CQM framework not only accelerates optimization but also produces physically credible and environmentally safer routes, highlighting its potential as a next-generation decision-support system for Arctic navigation under dynamic sea-ice conditions.}

\revise{
The remainder of this paper is organized as follows. 
Section~\ref{sec:related_work} reviews prior studies on Arctic maritime routing and hybrid quantum annealing optimization. 
Section~\ref{sec:background} introduces the theoretical background, summarizing the environmental datasets, the H3 hexagonal discretization framework, and the mathematical formulation of MIQP and CQM models. 
Section~\ref{sec:methodology} details the proposed method, including the H3-based ocean hexagonalization, feature-to-constraint mapping, and optimization model construction. 
Section~\ref{sec:exp_design} describes the experimental design and setup, outlining the computational environment, benchmark formulation, and solver configuration used for both synthetic and Arctic datasets. 
Section~\ref{sec:results} presents the comprehensive results and discussion, solver convergence and optimality analysis, scalability evaluation, real Arctic routing experiments, runtime budget sensitivity, and overall comparative insights between classical and hybrid quantum solvers. 
Finally, Section~\ref{sec:conclusion} concludes with key findings, current limitations, and prospective research directions, including the integration of POLARIS-compliant safety layers and real-time sea-ice forecasting capabilities.
}

\section{Related Works}
\label{sec:related_work}

Arctic navigation research has evolved rapidly in parallel with the growing accessibility of high-latitude shipping lanes.
Recent climatological projections indicate that navigable months along the NSR and TSR will continue to expand through mid-century, supporting potential year-round navigation under moderate emission scenarios \cite{zhang2023changing,Wang2024_YearRound}.
Environmental and community-oriented works further highlight the ecological risks of Arctic shipping, including the spread of invasive species via ballast discharge, and the limited participation of indigenous communities in governance~\cite{Saebi2025_Ballast}.
While these studies underscore the strategic value of Arctic shipping, they also highlight the persistence of non-linear sea-ice variability, which complicates route planning and makes optimization frameworks indispensable.
Earlier maritime routing approaches have primarily adopted deterministic or heuristic algorithms such as \revise{Dijkstra~\cite{choic2023} and A*~\cite{Kotovirta2023_Review,CHEN2025120956}}, which minimize travel distance or static environmental penalties.
Although efficient, these models fail to account for the coupling among ice thickness, drift, and navigational geometry, producing routes that may be locally feasible but physically unrealistic.
% \revise{Li \textit{et al.}~\cite{li2024feasibility}} demonstrated that navigable windows and voyage risk along the NSR are susceptible to moderate variations in ice thickness, emphasizing the importance of dynamic optimization using real-time data assimilation.
The demonstration of navigable windows and voyage risk along the NSR is sensitive to moderate variations in ice thickness, underscoring the importance of dynamic optimization with real-time data assimilation~\cite{li2024feasibility}.

Similarly, \revise{reviews from~\cite{Theocharis2023_Frontiers} and zone-based~\cite{song2022routeview}} proposed data-driven sea-ice routing methods, but their models relied on linearized costs and did not enforce curvature continuity, leading to discontinuous paths under changing sea-ice conditions.
Quadratic formulations, such as MIQP, have proven effective in related fields where pairwise interactions or curvature constraints are critical.

Recent benchmarking~\cite{Quinton2025_QAReview} studies have evaluated D-Wave’s Advantage annealer and LeapHybridCQMSolver against classical solvers such as Gurobi, CPLEX, and IPOPT across BLP, BQP, and MILP tasks. These works, using QUBO and Ising formulations with penalty-based constraints, show that quantum annealing can efficiently escape local minima and achieve near-optimal solutions on large quadratic problems, particularly for BQP. In contrast, classical solvers scale poorly under dense quadratic coupling. However, performance on MILP remains inconsistent due to sensitivity to penalty weights and decomposition overhead. These findings indicate that hybrid quantum annealing is promising for NP-hard routing problems involving quadratic curvature terms, such as Arctic ship routing, where scalability challenges motivate hybrid quantum-classical formulations.

In vehicle routing and energy flow optimization, quadratic terms have been used to capture smoothness, turning penalties, and non-linear flow consistency~\cite{Beyaztas2025,ROGNE2025121233,woodruff2019dynamic}.
These techniques provide conceptual and mathematical motivation for adopting MIQP in Arctic routing, where adjacent H3 cells represent spatially coupled environmental transitions rather than independent route segments.
Compared to terrestrial applications, however, the Arctic domain introduces additional non-convexity from dynamic sea-ice motion, requiring hybrid solvers capable of managing dense quadratic coupling under uncertain physical conditions.

The choice of spatial discretization has a significant impact on both computational performance and geophysical accuracy.
Traditional latitude-longitude grids introduce distortion near the poles and adjacency discontinuities across the 180° meridian, making them unsuitable for large-scale Arctic graph construction.y
In contrast, hexagonal Discrete Global Grid Systems, such as Uber’s H3~\cite{Uber2018H3}, provide equal-area tessellation, hierarchical scalability, and uniform neighborhood relationships~\cite{XIN2025104660,bousquin2021discrete}.
Recent work by Spadon~\cite{spadon2025goal} demonstrated that H3-based hexagonal grids outperform square tiling in maintaining spatial continuity and isotropy for navigation and reinforcement learning tasks.
The hexagonal uniformity eliminates the diagonal and edge ambiguities inherent in square grids, thereby~\cite{Vaidheeswaran2025HexGridRL} simplifying routing and spatial analysis through a consistent step cost, which is an essential property for further use in our approach.

\begin{figure}
\centering
\includegraphics[width=\columnwidth]{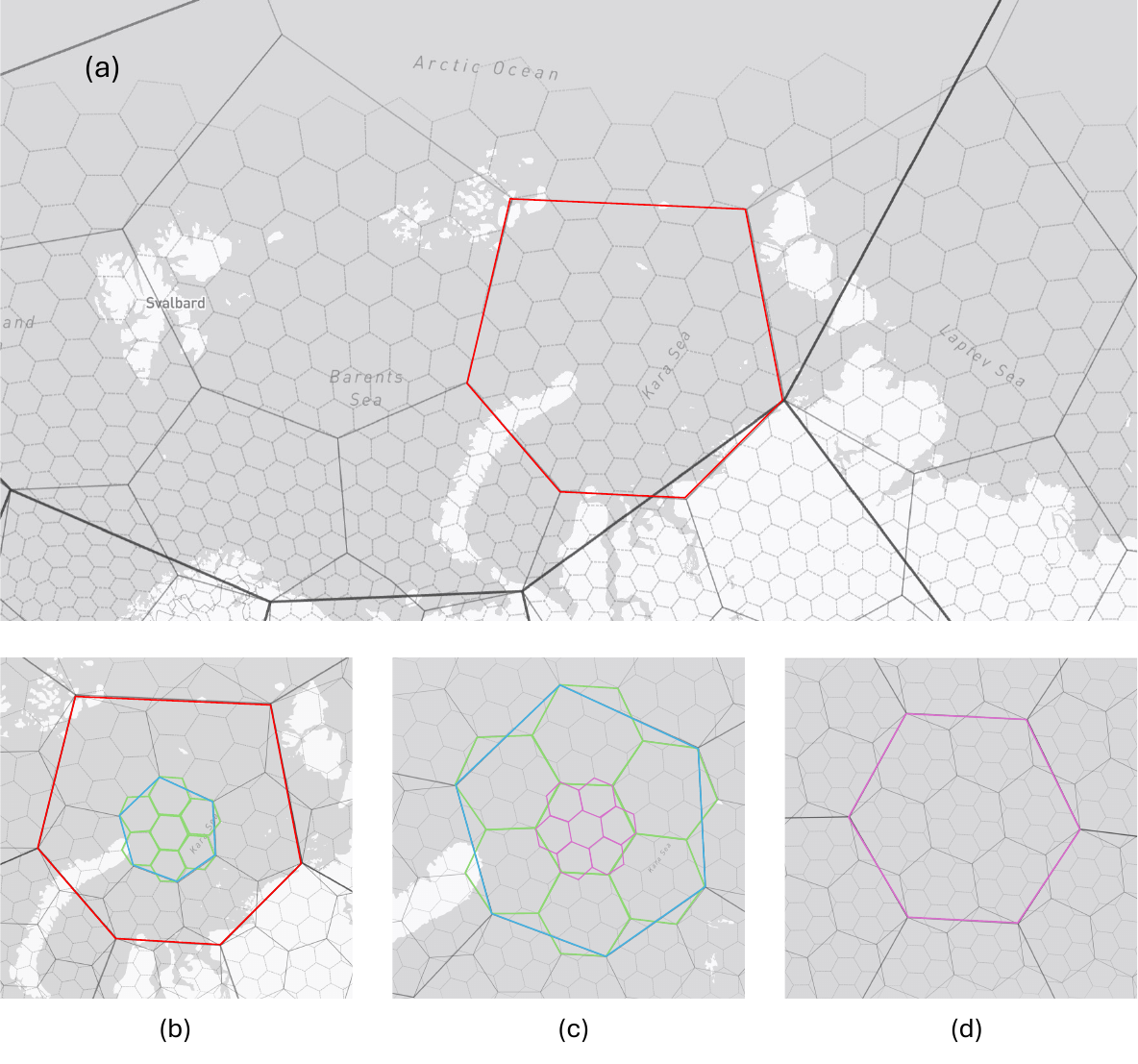}
\caption{Scalability of the H3 hierarchical grid applied to the Arctic Ocean. 
(a)~Global context highlighting the Barents and Kara Seas region. 
(b-d)~\newrevise{progressive refinement from coarse H3 resolution 3 to fine H3 resolution 6}. 
The nested structure maintains hexagonal adjacency and spatial continuity across levels, enabling scalable integration of sea-ice and oceanographic datasets for Arctic routing.}
\label{fig:scalability_of_hexagons}
\end{figure}

\begin{figure}
\centering
\includegraphics[width=\columnwidth]{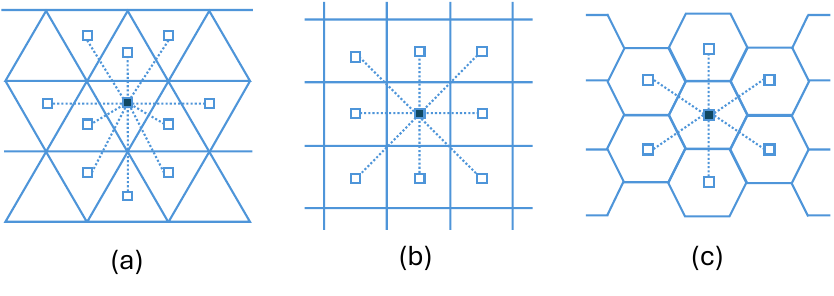}
\caption{Comparison of spatial discretization schemes: (a) triangular, (b) square, and (c) hexagonal (H3-based) lattices. The hexagonal configuration preserves isotropy and eliminates diagonal edge ambiguity, ensuring uniform neighbor connectivity and consistent step costs, which are essential for Arctic routing and spatial analysis. Figure adapted from~\cite{Vaidheeswaran2025HexGridRL}.}
\label{fig:hexagon_to_neighbors}
\end{figure}

To further illustrate H3's scalability and hierarchical consistency, Fig.~\ref{fig:scalability_of_hexagons} demonstrates the multi-resolution decomposition of the Arctic Ocean surface around the Barents and Karas Seas.
\newrevise{Finer H3 resolutions correspond to smaller hexagonal cells, meaning each cell covers a reduced geographic area. This increase in spatial granularity allows the grid to capture more localized sea-ice structures and environmental gradients that would be smoothed out at coarser levels. Such scalability is crucial for balancing computational efficiency with spatial precision in route optimization, as higher resolutions (e.g., 6 or 7) resolve detailed ice variability while still maintaining global consistency.}
% The grid hierarchy preserves geodisc continuity across resolutions, which allows seamless transitions between coars and fine spatial scales. This scalability is crucial for balancing computational efficiency with spatial precision in route optimization, as finer resolutions (e.g., 6 or 7) capture more localized sea-ice variations while maintaining global consistency.
These properties, as shown in Fig.~\ref{fig:hexagon_to_neighbors}, make H3 a natural fit for modeling Arctic ocean surfaces, enabling consistent neighbor connectivity and reducing topological artifacts in cost propagation.

Despite these advances, no prior study has unified (i) high-resolution H3 discretization, (ii) MIQP-based environmental optimization, and (iii) hybrid quantum annealing CQM solver benchmarking within a single computational framework \newrevise{for the Arctic ocean domain.}
This work fills that gap by demonstrating a physics-informed, multi-criteria Arctic routing model that leverages hexagonal graph topology and hybrid optimization.
Through comparative benchmarking against classical solvers (Gurobi, CPLEX), we quantify the performance and scalability of the D-Wave CQM approach under dense quadratic-term conditions, illustrating its promise for next-generation Arctic navigation systems.

\section{\revise{Background}}
\label{sec:background}
This section provides the theoretical and methodological background underpinning the proposed hybrid quantum annealing framework for Arctic route optimization. It introduces the environmental datasets, spatial discretization technique, and optimization formulations that collectively support the hybrid CQM implementation. This part provides an overview of the core components and their relationships, preparing the ground for the subsequent methodology.
% because it is the responsiblity as

\subsection{Sea-Ice and Environmental Data Modeling}
The navigability of Arctic sea routes is strongly governed by sea-ice dynamics and thermodynamic processes. Reliable representation of these processes requires integration of high-resolution environmental data.
This study uses the CMEMS ocean product Global Ocean Physics Analysis and Forecast~\cite{CMEMS_MDS_00016}, which provides gridded sea-ice variables with a horizontal resolution of approximately $0.083^{\circ}$ and daily temporal frequency. The following key parameters are incorporated: sea-ice thickness (\texttt{sithick}, vertical depth of sea ice indicating mechanical resistance), sea-ice age (\texttt{siage}, temporal maturity affecting hardness and melt behavior), eastward and northward ice-velocity components (\texttt{usi}, \texttt{vsi}, describing horizontal drift), ice concentration (\texttt{siconc}, fractional coverage of sea ice), and snow thickness over sea ice (\texttt{sisnthick}, accumulated snow depth influencing insulation and surface thermodynamics).

These variables collectively capture both the dynamic and thermodynamic states of the sea-ice environment and are directly linked to vessel operability and a safer route. Ice thickness and concentration determine the physical resistance along the route, while drift velocities define the directional bias induced by moving ice fields. By integrating these parameters at the level of spatial cells, the optimization model is able to represent navigability with improved physical consistency.

\subsection{Hexagonal Discrete Global Grid Systems (H3)}
Spatial discretization plays a crucial role in Arctic route modeling, as conventional latitude-longitude grids suffer from polar convergence and discontinuities at the antimeridian. To overcome these issues, this work employs the \textit{H3} hexagonal hierarchical grid developed by Uber Technologies~\cite{Uber2018H3}. The H3 framework partitions the Earth’s surface into nearly equal-area hexagons across multiple resolutions, providing isotropic adjacency and uniform step costs. 

The hexagonal configuration eliminates the diagonal and edge ambiguities inherent in square tiling and ensures seamless connectivity across the 180° meridian. Each H3 cell represents a fixed-area unit (approximately $250~\text{km}^2$ at resolution~5 in this study), from which environmental attributes are aggregated from the CMEMS dataset. These properties make H3 particularly suited for Arctic routing, where topological continuity and neighborhood uniformity are essential for stable cost propagation and curvature analysis.

\subsection{Mixed-Integer Quadratic Programming}
The routing problem can be formally expressed as an MIQP problem, in which linear terms represent distance and environmental penalties, and quadratic terms capture geometric smoothness and structural consistency. The general MIQP form~\cite{7526551} is written as:
\begin{equation}
\min_{x} \; \frac{1}{2} x^{\mathsf{T}} Q x + c^{\mathsf{T}} x \quad 
\text{s.t.} \; A x = b, \; Gx \leq h, \; x_i \in \{0,1\}
\end{equation}
where $x$ denotes the binary decision variables corresponding to route edges, $Q$ encodes quadratic curvature or turning penalties, and $c$ represents linear traversal costs incorporating sea-ice risk and geodesic distance. \newrevise{The superscript 
$\mathsf{T}$ denotes matrix transpose, ensuring that $x^{\mathsf{T}} Q x$ and 
$c^{\mathsf{T}} x$ follow standard quadratic and linear forms.} Constraints $A x = b$ and $Gx \leq h$ enforce flow continuity and structural feasibility. 

MIQP formulations are potent for modeling Arctic routing because they can express non-linear resistance, curvature continuity, and environmental trade-offs within a single optimization framework. However, as the quadratic term density increases, these problems become NP-hard, and classical solvers (e.g., Gurobi, CPLEX) experience exponential growth in runtime and memory usage. This motivates the exploration of hybrid quantum annealing methods that can handle dense quadratic coupling more efficiently.

\subsection{Constrained Quadratic Model and Hybrid Quantum Annealing Solvers}
The CQM formulation, implemented via D-Wave’s hybrid solver, extends the classical MIQP framework by embedding both equality and inequality constraints directly into the optimization model without converting them into penalties, as required in traditional QUBO-based approaches. The hybrid CQM solver~\cite{DWave_HybridSolvers_2022} operates on a quantum-classical architecture that combines the interpretability of constraint-based mathematical programming with the scalability of quantum annealing heuristics.

A quantum annealing approach refers to algorithms that emulate principles of quantum computation, such as energy minimization and probabilistic exploration on classical or hybrid hardware. The D-Wave CQM leverages these principles by orchestrating quantum subroutines with classical post-processing, enabling rapid convergence to near-optimal feasible solutions even for dense non-convex problems. 

This study adopts the CQM paradigm as a bridge between conventional mathematical optimization and emerging quantum computation. It enables direct comparison with classical solvers while demonstrating how hybrid quantum annealing mechanisms can deliver runtime and scalability advantages in large-scale Arctic route optimization tasks.

\section{Methodology}
\label{sec:methodology}
\begin{figure*}[!t]
\centering
\includegraphics[width=2.05\columnwidth]{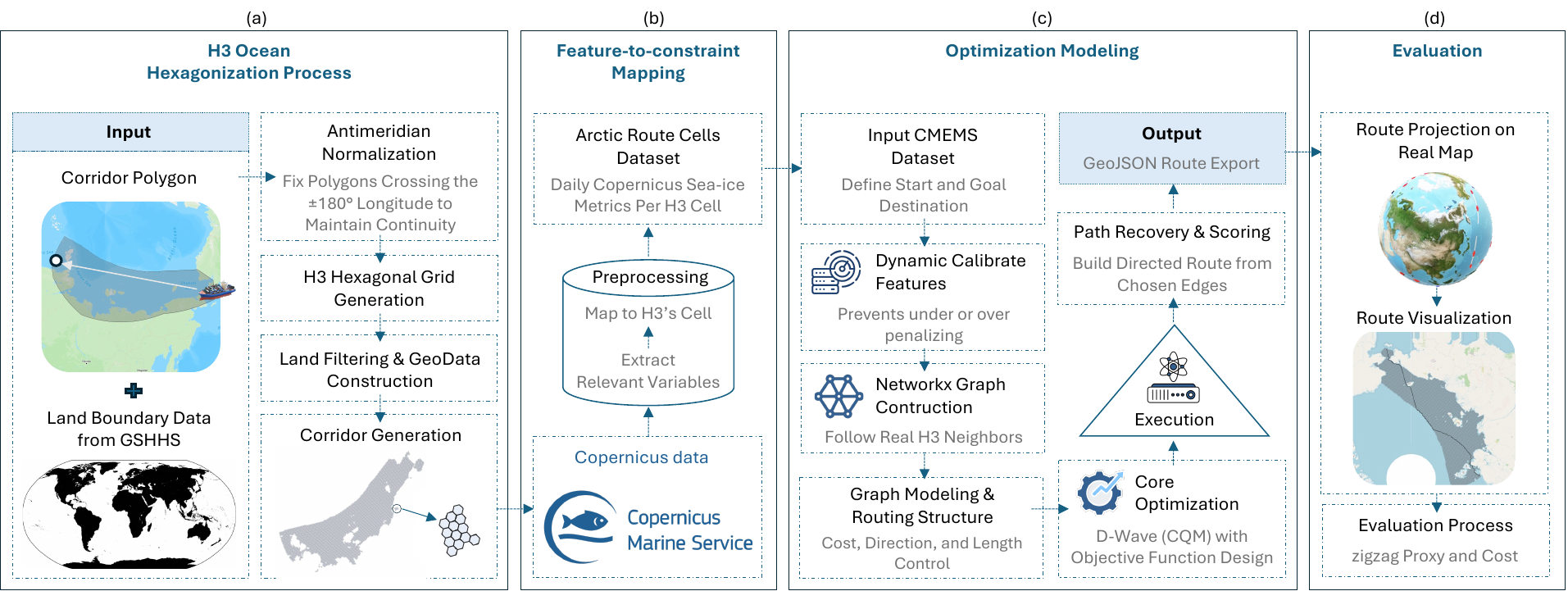}
\caption{\newrevise{End-to-end workflow of the Arctic ship routing framework. 
(a) H3 ocean hexagonization process, including corridor polygon input, antimeridian normalization, 
H3 grid generation and land filtering using the Global Self-consistent, Hierarchical, High-Resolution Geography (GSHHS) dataset. 
(b) Feature-to-constraint mapping, where CMEMS sea-ice variables are mapped to H3 cells through preprocessing and variable extraction. 
(c) Optimization modeling pipeline, consisting of network graph construction, objective design, and hybrid quantum execution via the D-Wave CQM solver, followed by path recovery and GeoJSON route export. 
(d) Evaluation stage, including route projection on a real map, visualization, and computation of evaluation metrics such as zigzag proxy and cost.}}
\label{fig_methodology}
\end{figure*}

\revise{Building upon the theoretical concepts introduced in Section~\ref{sec:background}, this section details the complete implementation pipeline of the proposed Arctic route optimization framework, including H3 ocean hexagonalization, feature-to-constraint mapping, and hybrid solver configuration.}
The overall workflow of the proposed Arctic route optimization framework is illustrated in Fig.~\ref{fig_methodology}. 
It consists of four primary stages: (1) H3 ocean hexagonalization, (2) feature-to-constraint mapping, (3) optimization modeling, and (4) evaluation. 
The design goal of this pipeline is to integrate high-resolution geospatial data from the CMEMS with a scalable, constraint-aware optimization framework capable of exploiting hybrid quantum-classical computation.

\subsection{H3 Ocean Hexagonalization Process}
The first stage of the proposed framework, corresponding to the H3 ocean hexagonalization process in Fig.~\ref{fig_methodology}\newrevise{(a)}, converts the Arctic maritime domain into a globally continuous and topologically consistent hexagonal lattice. This stage comprises four sequential components: corridor polygon, antimeridian normalization, H3 hexagonal grid generation, and land filtering \& GeoData construction, which collectively establish the spatial foundation for subsequent feature mapping and optimization.

\begin{figure*}[!t]
\centering
\includegraphics[width=2\columnwidth]{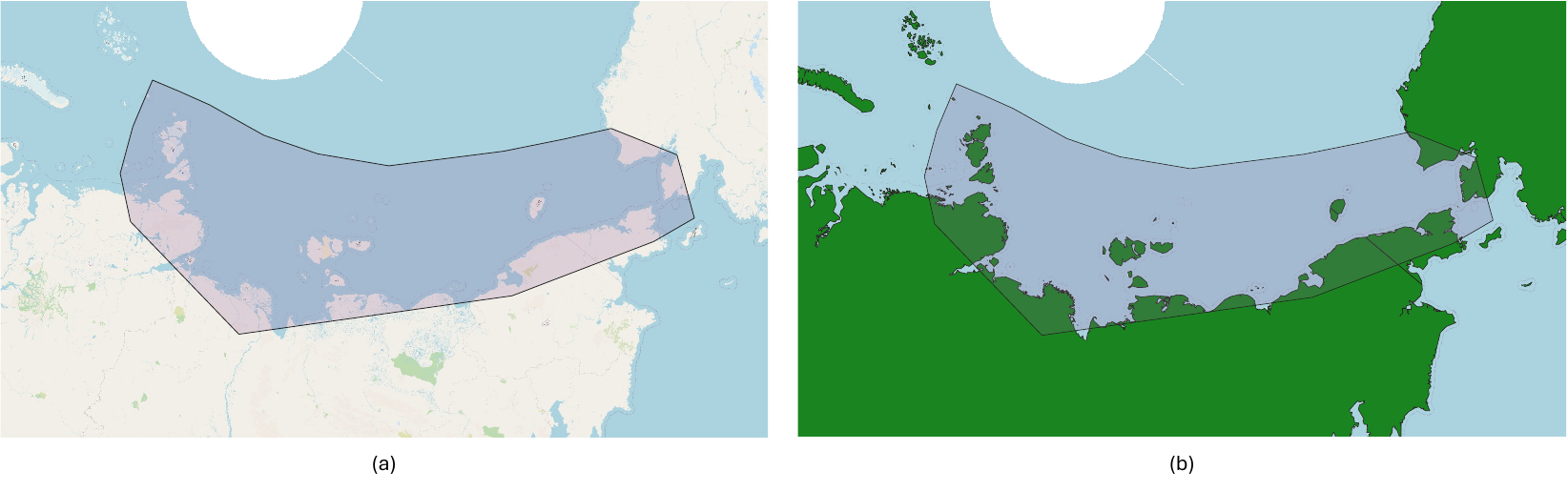}
\caption{Overview of the implemented H3 ocean hexagonalization and land-sea masking process.
(a) Integration of area of interest (AOI) boundary extraction, GSHHS shoreline masking, and H3 hexagonal network generation.
(b) Land-sea differentiation using the GSHHS dataset, where green regions indicate continental and island landmasses, and blue-shaded areas denote ocean surfaces. This vector-based representation enables accurate separation between terrestrial and marine domains, ensuring that only oceanic regions are retained for subsequent spatial analysis and route optimization.}
\label{fig:merge_gshhs_aoi}
\end{figure*}

The navigable Arctic corridor is first delineated as a polygonal boundary, which is defined in GeoJSON format under the WGS84 reference frame and represents the area of interest (AOI) encompassing the NSR and TSR sectors, as illustrated in Fig.~\ref{fig:merge_gshhs_aoi}(a).

As the AOI intersects the $\pm180^{\circ}$ meridian, a longitude-shift normalization is applied to maintain spatial continuity. For each vertex longitude $x_i$, the normalization operator $S_{\lambda}$ maps coordinates from $[-180^{\circ},180^{\circ}]$ to $[0^{\circ},360^{\circ}]$:
\[
S_{\lambda}(x_i)=
\begin{cases}
x_i+360^{\circ}, & x_i<0^{\circ},\\
x_i, & \text{otherwise}.
\end{cases}
\]
After geometric processing, the inverse operator $S_{\lambda}^{-1}$ restores the coordinates to the standard geographic range, guaranteeing polygonal continuity and consistency under the WGS84 (EPSG:4326) reference frame, \newrevise{the standard geographic coordinate reference system based on latitude and longitude.}.

The normalized polygon is discretized into equal-area cells using the H3 hierarchical spatial indexing system~\cite{Uber2018H3}. Each H3 cell provides uniform neighbor relationships and isotropic step costs. This geodesic configuration eliminates the polar distortion and edge discontinuities inherent in latitude-longitude grids, ensuring continuous adjacency and curvature analysis across the Arctic basin.
% (see Fig.~\ref{fig:merge_gshhs_aoi}(a)).

\begin{figure}
\centering
\includegraphics[width=\columnwidth]{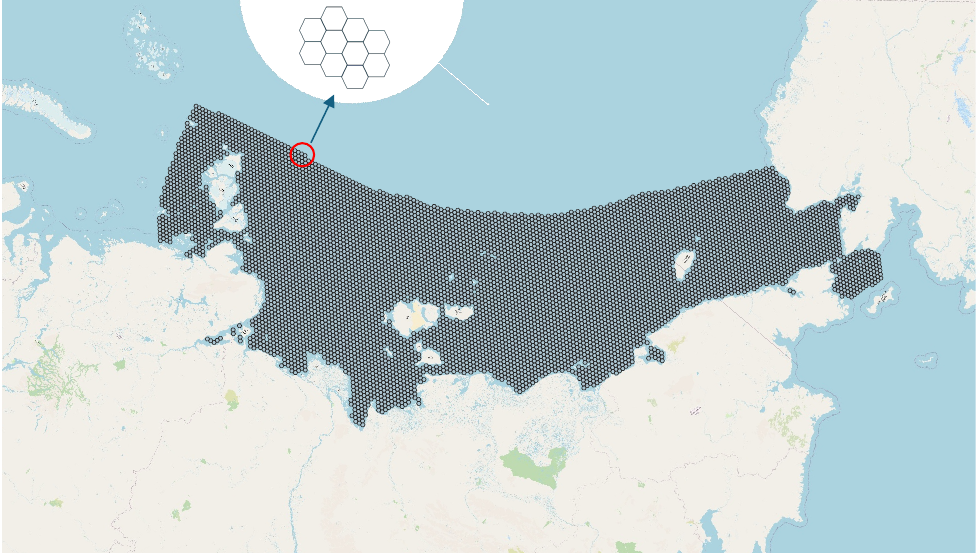}
\caption{Filtered hexagonal corridors produced after masking with GSHHS vectors. Each black hexagon represents a navigable ocean H3 cell retained for subsequent feature extraction and optimization.}
\label{fig:filter_land_corridor}
\end{figure}

To ensure that only oceanic regions are retained for routing, continental and island landmasses are removed using the GSHHS dataset~\cite{GSHHS_dataset}. The shoreline vectors are overlaid on the Arctic corridor polygon, as illustrated in Fig.~\ref{fig:merge_gshhs_aoi}(b), to distinguish terrestrial boundaries from marine surfaces. Any hexagonal cells intersecting the continental or island polygons are systematically excluded, leaving only oceanic areas suitable for navigation. The resulting filtered representation forms a continuous navigable corridor that preserves adjacency across the 180° meridian and delineates both open-water and near-coastal~\newrevise{zones as shown in Fig.~\ref{fig:filter_land_corridor}}. This spatial layer provides a geodesically consistent foundation for the next stage.

\subsection{Feature-to-Constraint Mapping}
% The second stage of the proposed framework integrates environmental observations from the CMEMS into the established hexagonal representation of the Arctic Ocean. 
% This step links each H3 hexagonal cell to the corresponding sea-ice variables derived from daily reanalysis and forecast data, thereby enriching the spatial domain with dynamic physical characteristics essential for subsequent optimization modeling (Fig.~\ref{fig:data_processing}).
The second stage of the workflow, corresponding to the feature-to-constraint mapping block in Fig.~\ref{fig_methodology}\newrevise{(b)}, integrates the CMEMS data into the hexagonal ocean grid. This process converts multiple sources of environmental observations into a unified Arctic route-cell dataset that quantitatively links sea-ice conditions to the spatial units used for route optimization.

\begin{figure}
\centering
\includegraphics[width=\columnwidth]{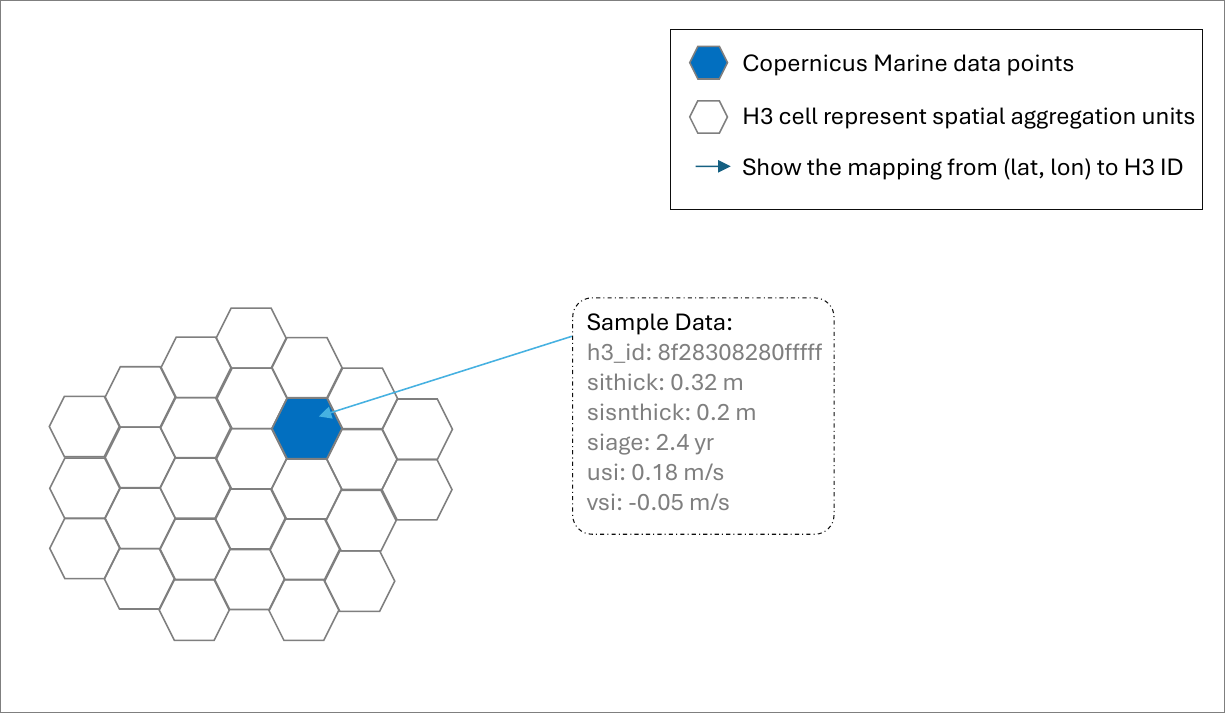}
\caption{Mapping of CMEMS sea-ice variables onto H3 hexagonal grids. Each dot represents a Copernicus Marine data point identified by latitude and longitude, mapped to a unique H3 cell via data aggregation. The sample cell displays its H3 ID and extracted attributes, including sea-ice thickness (\newrevise{\texttt{sithick}}), snow thickness over sea-ice (\texttt{sisnthick}), ice age (\texttt{siage}), sea-ice concentration (\texttt{siconc}), and eastward/northward ice velocities (\texttt{usi, vsi}). This step standardizes geospatial variables into uniform hexagonal units for further analysis and optimization.}
\label{fig:data_processing}
\end{figure}

Daily CMEMS sea-ice fields are loaded from the Global Ocean Physics Analysis and Forecast product~\cite{CMEMS_MDS_00016}.   
For each daily record, these variables are extracted from the NetCDF files and spatially aligned to a standard time index.
Each CMEMS grid point is then projected onto its corresponding H3 cell based on centroid coordinates $(\phi_i,\lambda_i)$ using nearest-neighbor geodesic interpolation.
This procedure aggregates environmental data within each hexagonal unit, ensuring a one-to-one correspondence between CMEMS observations and H3 indices.
Fig.~\ref{fig:data_processing} illustrates this mapping, where each dot represents a CMEMS data point linked to a unique H3 identifier id and its associated attributes.

The mapping produces a structured dataset in which every H3 cell stores averaged daily values of all sea-ice variables:
\[
F = \{f_{i,j}(t)\}, \quad i=1,\dots,n; \; j=1,\dots,m; \; t=1,\dots,T,
\]
where $f_{i,j}(t)$ denotes the value of variable $j$ at hexagon $i$ and time $t$. This matrix captures the spatial and temporal variability of sea-ice dynamics throughout the Arctic corridor.

The output of this stage is the Arctic route cells dataset, which is a geospatial data structure linking each navigable H3 cell to its CMEMS-derived environmental metrics. This dataset serves as the principal input to the subsequent optimization modeling stage, in which environmental information is translated into navigational costs and operational constraints.

\subsection{Optimization Modeling}
The third stage of the workflow, corresponding to the optimization modeling block in Fig.~\ref{fig_methodology}\newrevise{(c)}, transforms the processed environmental dataset into a constrained quadratic optimization problem.
This stage connects geospatial and physical parameters to the mathematical structure required for route computation.
The modeling pipeline proceeds through five main components: defining the start and goal destination, dynamically calibrating features, graph modeling routing structure, cost, direction, and length control, and D-Wave CQM with objective function design.

The routing problem begins with the selection of two H3 cells representing the departure ($s$) and destination \revise{($g$)} points within the navigable Arctic corridor. These cells serve as boundary nodes for the optimization process, anchoring the start and goal of the final route sequence. 
All subsequent graph construction and constraint enforcement operate with respect to these two terminals.

Each environmental attribute derived from the CMEMS dataset, including sea-ice thickness, age, concentration, snow depth, and velocity, is normalized using adaptive threshold values computed from the corridor dataset. 
Let \revise{$\tau_i, a_i, c_i, d_i$} denote the local sea-ice thickness, age, concentration, and snow depth at node $v_i$. 
For an adjacent pair $(i,j)$, we aggregate to a worst-case neighbor value:
\[
\begin{aligned}
\tau_{ij} &= \max(\tau_i, \tau_j), \quad\quad a_{ij} = \max(a_i, a_j),\\[4pt]
c_{ij} &= \min(c_i, c_j), \quad\quad d_{ij} = \max(d_i, d_j),
\end{aligned}
\]
with dataset-derived warning thresholds \\ 
\(\big(s_{\text{warn}}^{\text{thick}}, s_{\text{warn}}^{\text{age}}, s_{\text{warn}}^{\text{conc}}, s_{\text{warn}}^{\text{snow}}\big)\)
and observed bounds \\
\(\big(\tau_{\max}, a_{\max}, c_{\min}, d_{\max}\big)\),
we define the dimensionless penalties as follows:

\[
p_{\text{thick},ij} =
\max\!\left(0,
\frac{\tau_{ij} - s_{\text{warn}}^{\text{thick}}}
{\tau_{\max} - s_{\text{warn}}^{\text{thick}}}
\right)
\]
\[
p_{\text{age},ij} =
\max\!\left(0,
\frac{a_{ij} - s_{\text{warn}}^{\text{age}}}
{a_{\max} - s_{\text{warn}}^{\text{age}}}
\right)
\]
\[
p_{\text{conc},ij} =
\max\!\left(0,
\frac{s_{\text{warn}}^{\text{conc}} - c_{ij}}
{s_{\text{warn}}^{\text{conc}} - c_{\min}}
\right)
\]
\[
p_{\text{snow},ij} =
\max\!\left(0,
\frac{d_{ij} - s_{\text{warn}}^{\text{snow}}}
{d_{\max} - s_{\text{warn}}^{\text{snow}}}
\right)
\]
These calibrated penalties are used directly in the linear edge-cost term below, ensuring temporal robustness and preventing under- or over-penalization across different CMEMS snapshots.

Following feature calibration, the navigable H3 cells form the vertex set $V$ of an undirected graph $G = (V, E)$, where each edge $(i,j) \in E$ connects adjacent hexagons. 
The traversal cost $c_{ij}$ for each edge is computed as a weighted combination of the environmental penalties and geometric alignment factors introduced in the previous section (see calibrated penalties $p_{\text{thick},ij}$, $p_{\text{age},ij}$, $p_{\text{conc},ij}$, and $p_{\text{snow},ij}$). 
These quantities already incorporate the local sea-ice thickness, age, concentration, and snow depth values \revise{$(\tau_i, a_i, c_i, d_i)$} aggregated via the worst-case neighbor operations defined above. 
Each component is scaled by its respective dataset, derived warning threshold $(s_{\text{warn}}^{\text{thick}}, s_{\text{warn}}^{\text{age}}, s_{\text{warn}}^{\text{conc}}, s_{\text{warn}}^{\text{snow}})$ to maintain consistency across varying CMEMS datasets.

Using the normalized penalties $\{p_{\text{thick},ij}, p_{\text{age},ij}, p_{\text{conc},ij},\\ p_{\text{snow},ij}\}$ obtained from the previous calibration step, the linear traversal cost for each edge $(i,j)$ is formulated as a weighted combination of environmental and geometric factors:
\begin{equation}
\label{eq:cost}
\begin{aligned}
c_{ij} &= K_{\text{safety}}\!\left(
W_{\text{thick}}p_{\text{thick},ij} +
W_{\text{age}}p_{\text{age},ij} +
W_{\text{conc}}p_{\text{conc},ij}
\right. \\[4pt]
&\quad \left.
+\, W_{\text{snow}}p_{\text{snow},ij}
\right)
+ W_{\text{side}}\sigma_{ij}
+ W_{\text{lat}}\lambda_{ij}
+ H,
\end{aligned}
\end{equation}
where $K_{\text{safety}}$ scales the overall environmental risk level, $\sigma_{ij}$ and $\lambda_{ij}$ denote the side and lateral alignment deviations relative to the great-circle axis between the start and goal points, and $H$ is a small constant ensuring connectivity across equal-cost arcs.  
This formulation balances navigational efficiency with environmental risk mitigation by integrating both physical sea-ice constraints and geometric steering costs.  

To maintain route smoothness, a quadratic turning penalty is applied between consecutive edges that share a common vertex. 
For two connected arcs $(i,j)$ and $(j,k)$, the angular deviation penalty is expressed as:
\[
\omega_{(ij),(jk)} = W_{\text{turn}}\!\left(1 - \cos\theta_{ijk}\right),
\]
where $\theta_{ijk}$ represents the turning angle at node $v_j$ and $W_{\text{turn}}$ controls the curvature weight.  
This quadratic term discourages abrupt changes in direction, promoting realistic vessel trajectories that are both energy-efficient and physically feasible under Arctic navigation conditions.
\newrevise{While the quadratic turning penalty reduces abrupt changes in direction, it cannot fully prevent solver-induced oscillatory paths. Near ice boundaries, where multiple directions have similar costs, the model may still generate unrealistic zigzag movements that real vessels cannot safely execute due to increased rudder activity, hydrodynamic drag, and ice-interaction loads. As shown in prior path-planning studies, smooth heading transitions are essential for stable and energy-efficient motion~\cite{s20061550}. The zigzag proxy, therefore, provides an additional smoothness control to ensure that the optimized route is physically navigable under Arctic conditions.}

All linear and quadratic components introduced above are integrated into a unified constrained quadratic formulation that minimizes total traversal energy while maintaining structural and flow continuity. 
The complete optimization objective is expressed as:
\begin{equation}
\label{eq:objective_function}
\begin{aligned}
\min_{x,f} \Bigg[ &
\underbrace{\sum_{(i,j)\in E} c_{ij}x_{ij}}_{\text{environmental and distance cost}}
+
\underbrace{\sum_{((i,j),(j,k))\in \mathcal{P}} 
\omega_{(ij),(jk)}x_{ij}x_{jk}}_{\text{curvature penalty}} \\[4pt]
&+\, \Phi_{\text{deg}} + \Phi_{\text{len}}
\Bigg],
\end{aligned}
\end{equation}
where $x_{ij}\!\in\!\{0,1\}$ represents the binary decision variable indicating whether edge $(i,j)$ is active, and $f_{ij}$ denotes the associated directed flow variable that ensures path continuity between the start node $s$ and goal node $t$.

Equation~\eqref{eq:objective_function} consists of four principal components. \newrevise{Although the zigzag proxy is evaluated later outside the objective function, its role is closely related to the curvature term in (3), since both address the geometric smoothness of the route.}  
The first term aggregates the linear traversal costs $c_{ij}$ combining environmental and geometric effects, while the second term introduces the quadratic curvature penalty $\omega_{(ij),(jk)} = W_{\text{turn}}\!\left(1 - \cos\theta_{ijk}\right)$, promoting smooth directional transitions.
\newrevise{}
The remaining two components, $\Phi_{\text{deg}}$ and $\Phi_{\text{len}}$, represent soft quadratic penalties that preserve structural connectivity and path-length feasibility.  
Specifically, $\Phi_{\text{deg}}$ enforces local degree consistency by constraining each vertex $v$ to maintain the expected number of active incident edges $d_v$-two for intermediate nodes and one for the start ($s$) and goal ($g$) nodes:
\[
\begin{aligned}
\Phi_{\text{deg}} &= W_{\text{deg}}\!\!\sum_{v\in V}\!\left(\!\sum_{(i,j)\in E: i=v} x_{ij} - d_v\!\right)^{2}, \\[6pt]
d_v &=
\begin{cases}
2, & v \in V\setminus\{s,g\},\\[4pt]
1, & v\in\{s,g\}.
\end{cases}
\end{aligned}
\]
penalizes deviations from the expected node degree, ensuring that intermediate nodes have precisely two active connections (one incoming and one outgoing), while the start and goal nodes maintain degree one.  
Similarly, the path-length regularization term:
\[
\begin{aligned}
\Phi_{\text{len}} = W_{\text{len}}\!\!\Big[
&\max\!\left(0,\,L_{\min}-\!\!\sum_{(i,j)\in E}x_{ij}\!\right)^{2} \\[4pt]
&+\,\max\!\left(0,\,\!\!\sum_{(i,j)\in E}x_{ij}-L_{\max}\!\right)^{2}
\Big]
\end{aligned}
\]
constrains the total number of selected edges within a feasible range $(L_{\min},L_{\max})$, preventing unrealistically short or excessively long routes.  
Together, these terms preserve geometric realism, physical continuity, and navigational efficiency across the optimized path.

To guarantee a continuous and unbranched route between the start \revise{($s$)} and goal \revise{($g$)} nodes, directed flow conservation constraints, inspired by~\cite{Krauss2020}, are applied to all vertices:
\begin{equation}
\label{eq:condition}
\begin{aligned}
\sum_{(i,j)\in A} f_{ij} - \sum_{(j,k)\in A} f_{jk} &=
\begin{cases}
1, & i=s,\\[4pt]    
-1, & i=g,\\[4pt]
0,  & \text{otherwise},
\end{cases} \\[6pt]
&\quad 0 \le f_{ij} \le x_{ij}, \; \forall (i,j)\in A.
\end{aligned}
\end{equation}
% Equation~\ref{eq:condition} enforces one unit of flow originating from the start node and terminating at the goal node, ensuring that all intermediate vertices maintain flow balance and the resulting route remains singly connected without detached subpaths.
\newrevise{Equation~\ref{eq:condition} not only enforces unit flow conservation but also determines the structure of the resulting route produced by the optimizer. By requiring a net outflow of one unit at the start node and a net inflow of one unit at the goal node, while forcing all intermediate nodes to maintain zero net flow, the model selects exactly one incoming and one outgoing active edge for every visited vertex.
This decision structure guarantees that the solver constructs a single continuous path from $s$ to $g$, rather than producing disconnected segments or multiple branching trails.
In other words, the flow variables $f_{ij}$ act as routing indicators that ensure the final activated edges $\{x_{ij}\}$ form a valid, singly connected ship route consistent with the topology imposed by the H3 graph.}

The entire formulation is implemented as a CQM and executed on D-Wave’s hybrid quantum-classical solver~\cite{DWave_HybridSolvers_2022}.  
During hybrid execution, classical preprocessing identifies constraint structures and variable bounds, while quantum subroutines explore low-energy feasible configurations. 
Post-processing refinement then selects the optimal feasible solution, enabling rapid convergence to near-optimal Arctic routes even for dense and high-dimensional corridor graphs.

During execution, the D-Wave hybrid engine orchestrates classical preprocessing, quantum subproblem sampling, and post-processing repair to achieve energy-minimized feasible routes. The output is a binary activation set $\{x_{ij}\}$ that specifies the selected navigational edges, which is subsequently passed to the path recovery and evaluation stage for topological reconstruction and route visualization.

After optimization, the solver outputs a binary set of activated edges $\{x_{ij}\}$ and a set of continuous flow variables $\{f_{ij}\}$ representing feasible connections between H3 cells. 
Because the optimization model encodes only edge activation rather than traversal order, the raw binary solution may yield an undirected subgraph. Therefore, a dedicated reconstruction stage is used to obtain an explicitly ordered sequence of H3 cells that forms a continuous, geographically coherent route from the start node $s$ to the goal node \revise{$g$}.

Although the optimized decisions approximately satisfy degree and flow constraints, numerical tolerances or early solver termination may occasionally produce minor artifacts such as small detached branches or isolated subgraphs. 
These artifacts do not indicate formulation failure but arise naturally from the inherent relaxation and finite precision of large-scale quadratic models. 
To guarantee a continuous navigational path, a deterministic reconstruction procedure is applied after optimization.

All edges with $x_{ij}>0.5$ are first extracted to form an activated subgraph $G^{*}\subseteq G$. 
If directed flow variables are available, the traversal order follows the direction of positive flow values $(f_{ij}>0)$. 
Otherwise, the algorithm performs a shortest-path traversal within $G^{*}$ using edge weights $c_{ij}$ to recover the minimal geodesic sequence linking the start node $s$ and goal node \revise{$g$.}

When minor artifacts such as detached subgraphs or small redundant branches are identified, a minimal-cost relinking step is performed using the original cost matrix to restore full topological continuity. 
This relinking process does not alter the solver’s optimized edge selections but supplements missing links with the lowest-cost feasible connections so that all active edges form a single connected route. 
Such post-optimization repair or projection strategies are widely recognized in MIQP and graph-based optimization for improving feasibility without changing the original objective formulation \citep{SalvagninRobertiFischetti2024, HojnyJoormannLuthenSchmidt2021, NaikBemporad2021, TangXuAmosKolter2024}. 
The resulting network therefore represents a complete, continuous, and reproducible path suitable for geographic visualization and subsequent quantitative evaluation.
\newrevise{This reconstruction step does not introduce any new H3 cells or additional 
traversal. It only orders the solver-activated edges into a valid $s$ to $g$ path. 
Thus, no heuristic connections are added, the route cost remains unchanged, and 
the number of heuristically inserted cells is zero in all cases.}

Each reconstructed node $v_i$ is then converted to geographic coordinates $(\phi_i,\lambda_i)$ corresponding to the centroid of its H3 cell. 
The ordered sequence $\{v_s,\ldots,v_t\}$ is projected onto the Earth's ellipsoidal surface to form a continuous polyline through great circle interpolation between consecutive nodes. 
The geodesic distance between two nodes is computed by the haversine relation:
\begin{equation}
d_{ij}=2R_{\oplus}\arcsin\!\left(
\sqrt{\sin^2\!\frac{\phi_j-\phi_i}{2}
+\cos\phi_i\cos\phi_j\sin^2\!\frac{\lambda_j-\lambda_i}{2}}
\right),
\label{eq:geodesic_route}
\end{equation}
where $R_{\oplus}$ is the mean Earth radius (6,371~km).  
This step transforms the optimized graph path into a geographic route that preserves the real-world spatial geometry of the Arctic maritime corridor.

\begin{figure*}
\centering
\includegraphics[width=2\columnwidth]{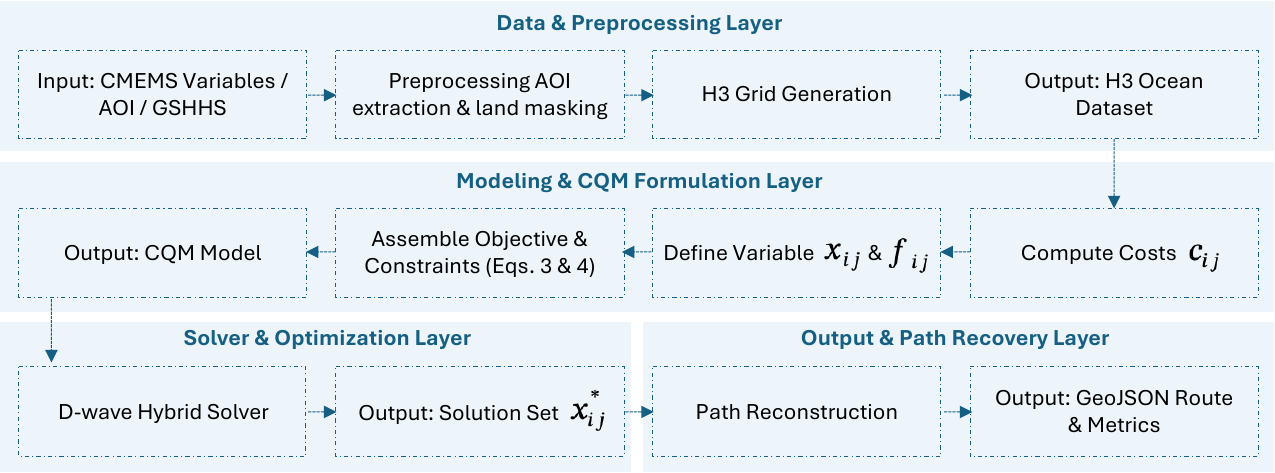}
\caption{Workflow of the proposed Arctic route optimization framework implemented on the D-Wave hybrid CQM solver. The four hierarchical layers, such as Data \& Preprocessing, Modeling \& CQM Formulation, Solver \& Optimization, and Output \& Path Recovery, illustrate the transformation from environmental data to optimized navigational routes. Each transition explicitly denotes the exchanged variables and intermediate outputs that maintain consistency between geophysical modeling, mathematical formulation, and solver execution.}
\label{fig:flow_diagram}
\end{figure*}

\revise{Fig.~\ref{fig:flow_diagram} illustrates the data-to-solver workflow for the proposed hybrid quantum annealing framework for Arctic route optimization.} 
The process consists of four sequential layers:

\begin{enumerate}
    \item \revise{Data \& Preprocessing Layer:}
    Multi-variable sea-ice inputs from CMEMS (thickness, age, drift, and concentration) are combined with AOI boundaries and the GSHHS shoreline dataset. 
    AOI extraction and land masking are applied, followed by H3 hexagonal grid generation to ensure global continuity across the antimeridian. 
    The resulting H3 Ocean Dataset provides uniform spatial cells forming the computational foundation for route modeling.

    \item \revise{Modeling \& CQM Formulation Layer:}
    The H3 dataset is transformed into a constrained optimization structure. 
    For each edge $(i,j)$, the traversal cost $c_{ij}$ is computed based on environmental penalties. 
    Binary and flow variables $(x_{ij}, f_{ij})$ represent route selection and directional continuity. 
    These are assembled into the Constrained Quadratic Model (CQM) through the objective and constraint equations (Eqs.~\eqref{eq:objective_function} and \eqref{eq:condition}), yielding the CQM model for solver execution.

    \item \revise{Solver \& Optimization Layer:}
    The formulated CQM is executed on D-Wave’s hybrid solver, which combines classical preprocessing and quantum annealing to efficiently minimize the objective energy. 
    The solver outputs a solution set $x_{ij}^*$ representing the optimized route edges that satisfy environmental and flow constraints.

    \item \revise{Output \& Path Recovery Layer:}
    The binary solution set $x_{ij}^*$ is post-processed to reconstruct a continuous navigable route connecting the start and goal nodes within the H3 grid. 
    The resulting path is exported as a GeoJSON route and evaluated using key metrics such as route length, zigzag proxy, and estimated CO$_2$ emissions, ensuring both computational efficiency and environmental realism.
\end{enumerate}

\section{\revise{Experimental Design and Setup}}
\label{sec:exp_design}
This section describes the overall experimental environment, benchmark formulation, and solver configurations used to evaluate the proposed optimization framework.

\subsection{Experimental \revise{Setup}}
All experiments were conducted on a local workstation (Apple Mac Studio) equipped with an Apple M2 Max chip and 64 GB of unified memory, 12 logical processors, and Python 3.10.  
Classical solvers were executed using Gurobi Optimizer v12.0.3 and IBM CPLEX v22.1.1.0, both operating in native CPU mode with default multithreading enabled, and a relative optimality gap tolerance of 0\% was applied, \newrevise{meaning the solver must prove that the best-found solution is globally optimal by matching the best bound}, to ensure near-optimal convergence within practical runtime limits.
The experiments utilized the D-Wave Ocean SDK v8.4.0, employing the LeapHybridCQMSampler interface for remote execution on D-Wave’s cloud-based CQM solver \revise{(Advantage2\_system1.6)}.  
All timing results reported in this study correspond to wall-clock runtimes measured on this configuration, including solver initialization and data-loading overhead.  
This setup ensures consistent benchmarking across solver paradigms, enabling reproducible comparative runtime analyses.

A concise summary of the computational and data configuration is provided in Table~\ref{tab:exp_setup}. The table lists the hardware platform, solver versions, CQM budgets, Arctic spatial grid, ocean–ice dataset, preprocessing routines, and routing domain used in all experiments.
\begin{table*}
\caption{\newrevise{Summary of the computational environment, solver configuration, spatial grid, and data sources used in the experimental setup.}}
\label{tab:exp_setup}
\centering
\begin{tabular*}{\tblwidth}{@{} LLL@{} }
\toprule
\textbf{Component} & \textbf{Settings} & \textbf{Source / Tool} \\
\midrule

Hardware
& Local server; D-Wave cloud hybrid backend 
& D-Wave Leap \\

Solvers
& Gurobi 12.0.3; CPLEX 22.1.1; D-Wave Hybrid CQM 
& Vendor implementations \\

CQM Budgets Analysis
& 30, 60, 90, 120, 150s 
& LeapHybridCQMSampler \\

Arctic Grid 
& H3 resolution 5 (2,045, 5,130, 7,884 nodes) 
& Uber H3 library~\cite{Uber2018H3} \\

Ocean–Ice Data
& Sea-ice thickness, age, velocity, concentration, snow thickness
& CMEMS Global Ocean Physics~\cite{CMEMS_MDS_00016} \\

Preprocessing 
& NetCDF/JSON parsing; H3 aggregation; adjacency construction 
& Python (NumPy, Pandas, h3) \\

Routing Corridor
& Bering Strait $\rightarrow$ Sabetta corridor (NSR region) 
& Geographic bounding box \\

\bottomrule
\end{tabular*}
\end{table*}

\subsection{Synthetic Benchmark Design}
Designing a synthetic benchmark that mirrors the full mathematical structure of the \newrevise{Arctic routing system~(ARS)} model is essential to isolate solver behavior from geophysical effects. 
The ARS formulation in Equation~\eqref{eq:objective_function} integrates traversal costs, curvature penalties, and soft structural constraints into a unified MIQP objective. 
To replicate this complexity in a controlled setting, we construct a structural analog benchmark maintaining equivalent algebraic coupling and constraint relaxation.
\begin{table*}[!t]
\caption{This table summarizes the structural correspondence between the real \newrevise{ARS} formulation and the synthetic MIQP benchmark, demonstrating that all core decision and cost components are preserved with minor relaxation in constraint logic.}
\label{tab:synth_comparison}
\centering
\begin{tabular*}{\tblwidth}{@{} LLLL@{} }
\toprule
\textbf{Component} & 
\textbf{ARS Formulation (Equation~\eqref{eq:objective_function})} & 
\textbf{Synthetic MIQP (Equations~\eqref{eq:synth_objective} \& \eqref{eq:synth_constraints})} & 
\textbf{Equivalence} \\
\midrule
Decision variables & Route-arc binary $x_{ij}\!\in\!\{0,1\}$ & Binary activation $x_i\!\in\!\{0,1\}$ & 1:1 \\
\midrule
Quadratic terms & $\omega_{(ij),(jk)}x_{ij}x_{jk}$ (curvature) & $Q_{ij}x_i x_j$ (random couplings) & Preserved \\
\midrule
Soft constraints & $\Phi_{\mathrm{deg}}\!+\!\Phi_{\mathrm{len}}$ (structural) & $\textit{slack\_penalty}(s_1+s_2)$ & Equivalent \\
\midrule
Constraint logic & Flow-continuity, start-goal & Cardinality and soft bounds & Relaxed analog \\
% \midrule
% Solver type & MIQP / CQM hybrid & MIQP / CQM hybrid & Identical \\
\midrule
Context & Geospatial routing & Abstract binary network & Abstracted \\
\midrule
Overall similarity & – & $\approx$80\% of ARS complexity & High \\
\bottomrule
\end{tabular*}
\end{table*}
\subsubsection{Formulation Rationale}
\label{subsec:synthetic}
The benchmark generalizes the ARS structure as:
\begin{equation}
\label{eq:synth_objective}
\begin{aligned}
\min_{x,s_1,s_2}\; f(x,s_1,s_2)
&= \sum_i c_i x_i
   + \sum_{i<j} Q_{ij}x_i x_j  \\
&\quad + \textit{slack\_penalty}\,(s_1+s_2),
\end{aligned}
\end{equation}
subject to the following constraints:
\begin{equation}
\label{eq:synth_constraints}
\begin{aligned}
&L \le \sum_i x_i \le U, \\[2pt]
&T_1 - \Big(\sum_i w_i x_i\Big)^2 \le s_1, \\[2pt]
&T_2 - \!\!\sum_{i<j}\! R_{ij}x_i x_j \le s_2.
\end{aligned}
\end{equation}
% Equation.~\eqref{eq:objective_function} and~\eqref{eq:synth_objective} share identical solver, theoretic class-nonconvex MIQP with binary decision variables, indefinite quadratic couplings, and soft relaxed constraints. Therefore, differing only by semantics of variable interpretation.
Here, $s_1$ and $s_2$ are non-negative \textit{slack variables} that soften otherwise rigid quadratic constraints, 
allowing small violations at a penalized cost. 
The first slack variable $s_1$ relaxes the weighted-sum constraint on the binary activations, 
while $s_2$ relaxes the pairwise coupling constraint.
This relaxation strategy, inspired by the soft quadratic formulation in~\cite{lwxq-4myj}, enables controlled constraint flexibility and tunable benchmark difficulty.
The scalar parameter \textit{slack\_penalty} controls the trade-off between exact feasibility and solver convergence. 
Equations~\eqref{eq:objective_function} and~\eqref{eq:synth_objective} share the same nonconvex MIQP structure with binary decision variables, indefinite quadratic couplings, and soft relaxed constraints, differing only by the semantics of variable interpretation.

\subsubsection{Structural Correspondence}
Table~\ref{tab:synth_comparison} summarizes the one-to-one mapping between the ARS and synthetic MIQP, preserving $\sim$80\% of algebraic complexity relevant to solver hardness.

\subsubsection{Solver Configuration and Protocol}
Each synthetic instance was solved using three solver paradigms:
\begin{enumerate}
\item Gurobi: establishes the reference optimum by proving global optimality~$f^*$;
\item CPLEX: provides independent verification through a distinct branch-and-cut heuristic;
\item CQM: performs quantum-classical heuristic search without a formal optimality proof but with rapid convergence to near-optimal feasible energies.
\end{enumerate}
All solvers were executed without explicit time limits, allowing each to terminate naturally, either upon proof of optimality (for classical solvers) or upon hybrid completion (for CQM).  
This ensures that recorded runtimes accurately reflect the intrinsic solver efficiency rather than user-imposed limits and provides an equitable basis for cross-paradigm comparison.

The following experiments are designed to evaluate both the computational behavior and the physical interpretability of the proposed framework. 
Specifically:
\begin{enumerate}
    \item Solver convergence and optimality: compares Gurobi, CPLEX, and D-Wave CQM on synthetic MIQP benchmarks to assess proof-based versus heuristic convergence speed.
    \item Scalability under quadratic growth: examines runtime and feasibility as the number of quadratic couplings increases, reflecting the rising dimensionality of Arctic graphs.
    \item Real Arctic routing evaluation: applies the model to CMEMS sea-ice datasets to validate route smoothness, energy efficiency, and CO\textsubscript{2} proxy improvements across spatial scales.
    \item Analysis for budget recommendation: analyzes solver stability under extended hybrid time limits to identify the most efficient configuration for practical deployment.
\end{enumerate}
Together, these experiments provide a comprehensive assessment of computational performance, scalability, and physical realism, setting the stage for the quantitative analyses presented in Section~\ref{sec:results}.

\begin{table*}[!t]
\caption{Performance comparison of Gurobi, CPLEX, and D-Wave CQM across increasing quadratic term densities. 
For the 1,867-term case, CPLEX failed to reach optimality within a reasonable time and remained unresolved at termination. 
Notably, CQM’s sub-optimal solutions are numerically equivalent to the optimal results of Gurobi and CPLEX, while achieving them within only a few seconds.}
\label{tab:solver_perf}
\centering
\begin{tabular*}{\tblwidth}{@{} CLLL@{} }
\toprule
\textbf{\makecell{Number of\\Quadratic Terms}} & \textbf{Solver} & \textbf{Best Objective Value} & \textbf{Time to Optimal (s)} \\
\midrule
\multirow{3}{*}{1,194} & Gurobi & -224.0030 & 56 \\
 & CPLEX & -224.0031 & 372.82 \\
 & \textbf{CQM} & \textbf{-224.0030} & \textbf{5} \\
\midrule
\multirow{3}{*}{1,316} & Gurobi & -208.6590 & 116 \\
 & CPLEX & -208.6590 & 2,682.18 \\
 & \textbf{CQM} & \textbf{-208.6590} & \textbf{5} \\
\midrule
\multirow{3}{*}{1,867} & Gurobi & -237.3735 & 5,425.154 \\
 & CPLEX & -137.8564 & 146,079.36 \\
 & \textbf{CQM} & \textbf{-237.3735} & \textbf{5} \\
\bottomrule
\end{tabular*}
\end{table*}

\section{\revise{Results and Discussion}}
\label{sec:results}
This section integrates quantitative results and interpretative analyses, progressing from synthetic solver benchmarking to realistic Arctic routing scenarios.  
The discussion emphasizes computational scaling, physical interpretability, and operational feasibility of the proposed hybrid optimization framework.

\subsection{Solver Convergence and Optimality}
\label{subsec:exp1}
\begin{figure}
\centering
\includegraphics[width=1.03\columnwidth]{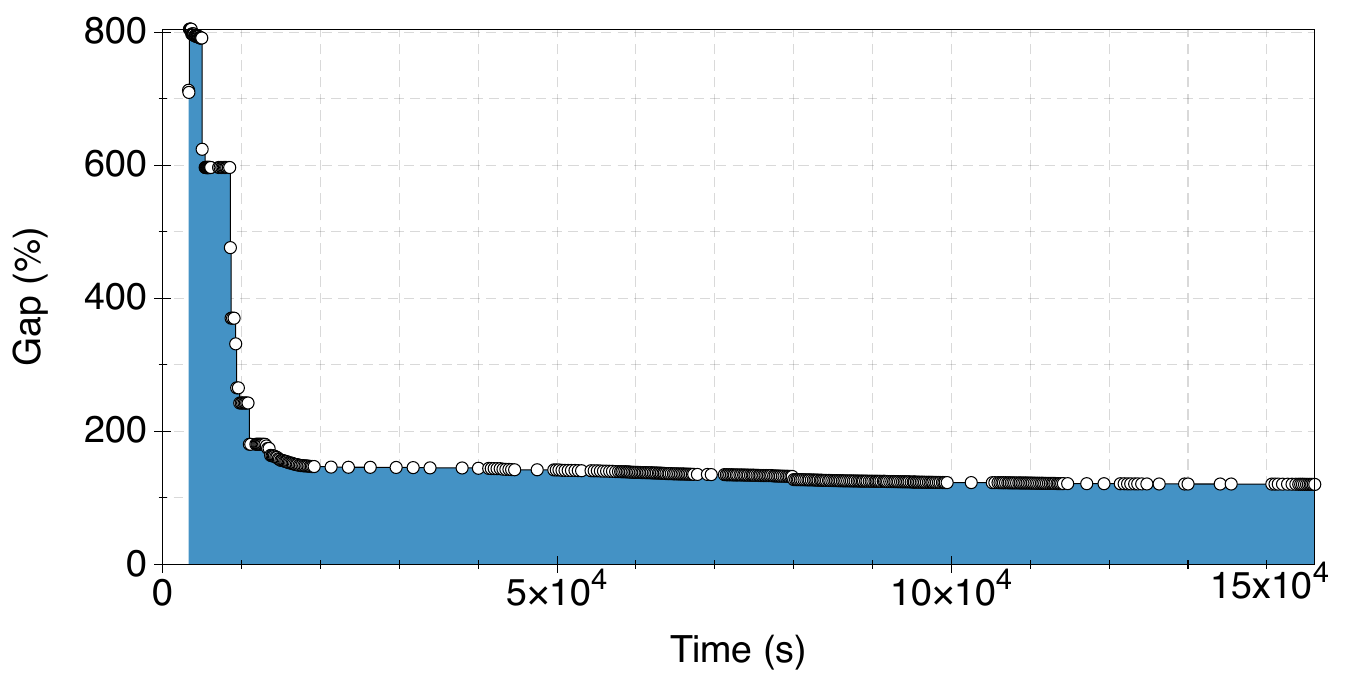}
\caption{CPLEX convergence trajectory for the 1,867-term case, showing the optimality gap decreasing rapidly before stagnating near zero. 
The near-zero gap indicates convergence to the true optimum, but proof was not completed within the allowed runtime (see the third row of Table~\ref{tab:solver_perf}).}
\label{fig:cplex_gap}
\end{figure}

We compare the \textit{time to optimality} for Gurobi and CPLEX against the time to best feasible energy for the D-Wave CQM on the synthetic MIQP benchmark described in Section~\ref{subsec:synthetic}. 
No explicit time budget is imposed: classical solvers are allowed to run until they either prove global optimality or reach prolonged stagnation, while the CQM executes until hybrid completion. 
Performance statistics are summarized in Table~\ref{tab:solver_perf}, and the CPLEX convergence trajectory for the largest case is shown in Fig.~\ref{fig:cplex_gap}.

As shown in Table~\ref{tab:solver_perf}, Gurobi and CPLEX converge to numerically identical global optima across all test sizes, validating the consistency of both classical solvers.  
However, the proof times differ significantly: Gurobi establishes optimality within minutes, whereas CPLEX requires up to two orders of magnitude longer, often exceeding one day for the largest instance.  
The D-Wave~CQM, by contrast, reaches equivalent objective values in roughly~5s without explicit proof, \newrevise{means that the hybrid CQM solver returns a high-quality feasible solution but does not provide a mathematical certificate 
of global optimality, unlike classical solvers}.  

Fig.~\ref{fig:cplex_gap} reveals how CPLEX’s optimality gap drops sharply at first but flattens near zero, implying the solver finds a strong incumbent early but struggles to tighten the dual bound.  
This plateau behavior is common in branch-and-cut algorithms for nonconvex MIQPs, where verifying global optimality becomes exponentially expensive once near-optimal solutions are found.  
The CQM’s ability to bypass this proof phase while still reaching equivalent energies demonstrates a key hybrid quantum annealing efficiency: practical optimality can be achieved much faster than formal verification.  
In large-scale engineering contexts, where decision quality outweighs mathematical proof, such near-optimal convergence within seconds is a tangible computational benefit.  
These results suggest an emerging hybrid workflow: use the CQM to obtain feasible high-quality routes rapidly, and subsequently employ Gurobi for proof-based validation when certification is required.

\subsection{Scalability under Quadratic Growth}
\label{subsec:exp2}

To evaluate how solver performance scales with increasing quadratic density, mimicking the growing complexity of high-resolution Arctic graphs.

\begin{figure}
\centering
\includegraphics[width=1.04\columnwidth]{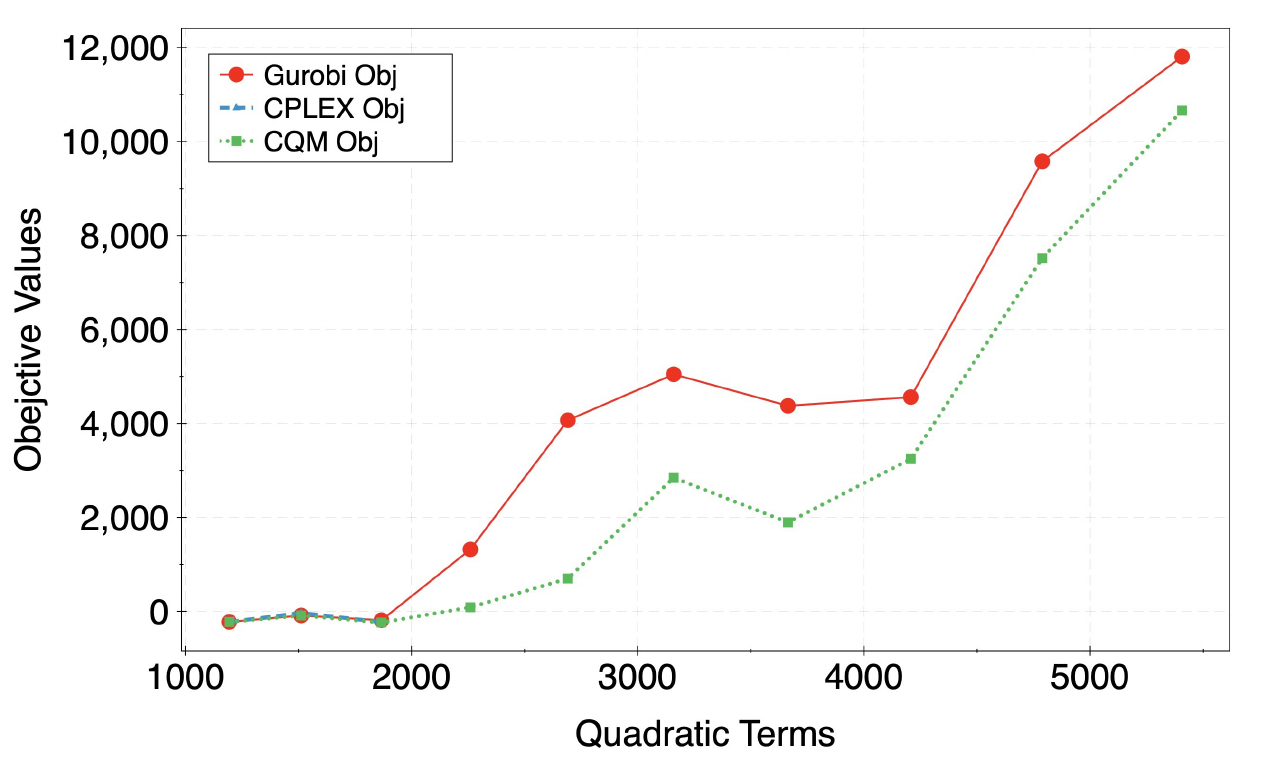}
\caption{Comparison of solver objective trends under capped time budgets. 
Gurobi was limited to 1,000s and CPLEX to 2,000s; even when extended to 10,000s, CPLEX remained infeasible beyond 1,867 quadratic terms. 
In contrast, the D-Wave CQM solver consistently produced feasible solutions within approximately 5s across all tested quadratic term densities, achieving lower (more optimal) objective values than the classical solvers.}
\label{fig:scalability_trend}
\end{figure}

Fig.~\ref{fig:scalability_trend} shows that while all solvers exhibit monotonic objective growth as the number of quadratic terms increases, divergence becomes evident once the problem surpasses $\sim$2,000 terms.  
Gurobi’s performance deteriorates as branching depth grows, while CPLEX begins producing infeasible or incomplete solutions beyond this threshold, even with extended runtimes.
By contrast, CQM maintains stable energy trajectories and completes each instance within a constant hybrid cycle of about~5s.

This behavior highlights the inherent scalability challenge of classical solvers: as quadratic coupling density increases, constraint matrices grow denser, and the branch-and-bound trees expand exponentially.  
Quantum annealing-based hybrids mitigate this effect by exploring the global energy surface through probabilistic tunneling and subproblem decomposition rather than exhaustive enumeration.  
The near-constant runtime observed for CQM across problem sizes indicates that the solver’s hybrid orchestration efficiently distributes workloads between classical preprocessing and quantum subroutine sampling.  
Such scalability is critical for Arctic routing, where each increase in H3 resolution introduces thousands of new variables and quadratic curvature terms.  
These findings confirm that the CQM can maintain feasibility and near-optimality for problem sizes where classical methods become impractical, marking a clear computational crossover point that is favorable to hybrid quantum approaches.

\subsection{Real Arctic Routing Evaluation}
\label{subsec:exp3}

To validate the proposed Arctic route optimization framework on real sea-ice and oceanographic conditions and to evaluate solver behavior, feasibility, runtime, and physical interpretability, under realistic multi-criteria constraints.

The experiments use the \textit{Global Ocean Physics Analysis and Forecast} product from the Copernicus Marine Environment Monitoring Service (CMEMS)~\cite{CMEMS_MDS_00016}.  
The selected snapshot corresponds to late-season Arctic conditions with a maximum observed sea-ice thickness of approximately~0.72m, representative of medium first-year ice.  
According to the navigability of NSR route~\cite{CHEN2022676}, medium first-year ice lies in the 0.7-1.2m range, marking the lower boundary of typical navigable windows.  
Prior studies have shown that voyage feasibility and vessel performance along the Northern Sea Route are highly sensitive to ice-thickness fluctuations within this interval~\cite{li2024feasibility}, while satellite analyses by~\cite{rs16162983} with recent thicknesses hovering around the 1.0-2.0m range, making thinner ice types dominant.  
Thus, this dataset offers a realistic benchmark for evaluating navigability and solver robustness under operational Arctic conditions.

Three solver paradigms, Gurobi, CPLEX, and D-Wave CQM, were applied to identical discretized corridors at different resolutions using the H3 grid.  
The small, medium, and large graphs comprise 2,045,~5,130, and 7,884 H3 nodes, respectively, with corresponding quadratic term counts of 5,596, 14,234, and 21,959.  
All solvers employed the same objective in Equation~\eqref{eq:objective_function}, integrating environmental penalties (thickness, age, concentration, snow depth) and geometric curvature terms.  
To ensure parity, Gurobi and CPLEX were capped at 4,200s, consistent with the scalability results in Section~\ref{subsec:exp2}.
% , whereas the D-Wave CQM completed each hybrid execution within 5-30 s depending on graph size.

\begin{figure*}
\centering
\includegraphics[width=2\columnwidth]{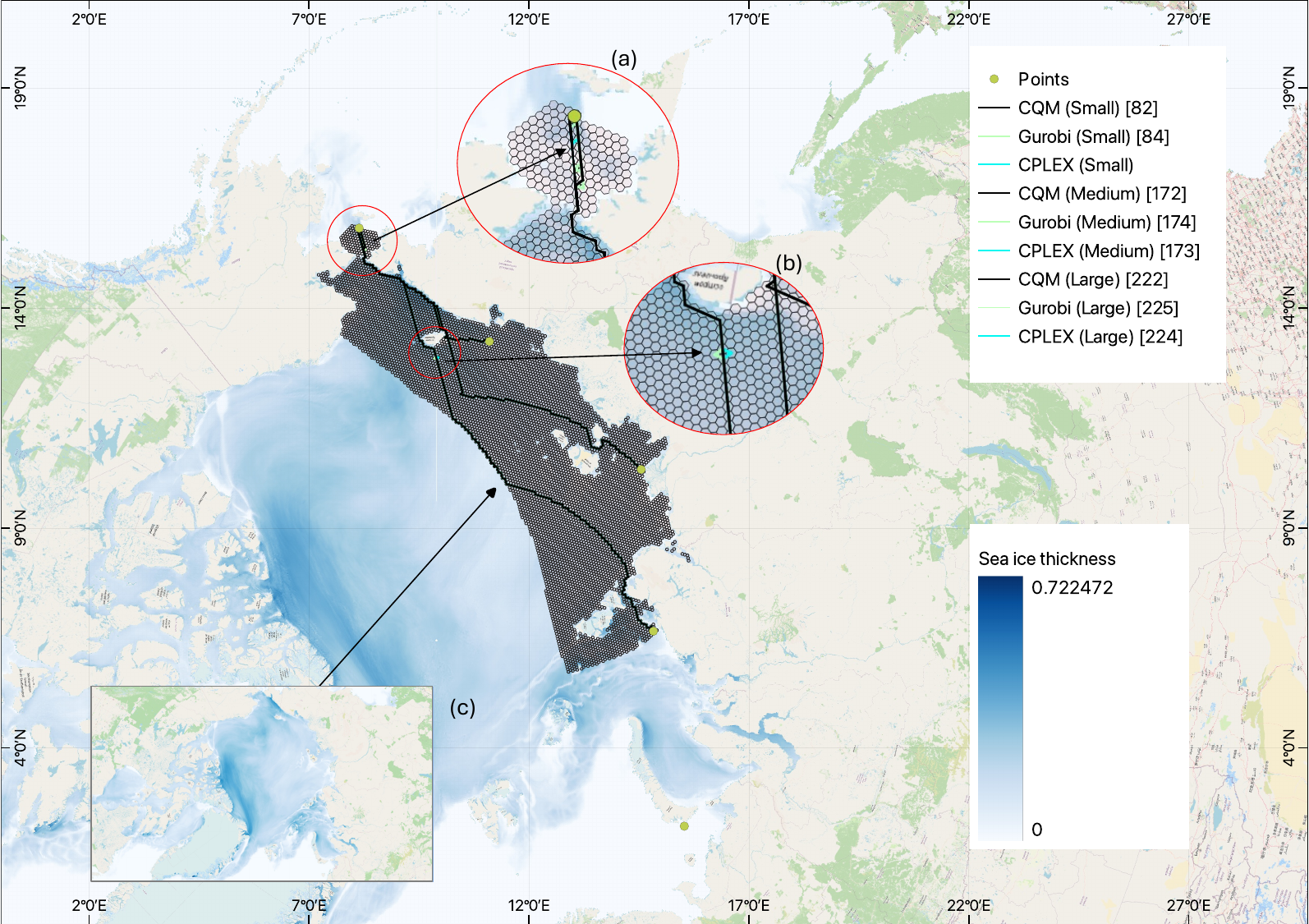}
\caption{Optimized Arctic routes across small, medium, and large problem scales, overlaid on sea-ice-thickness data. The visualized paths correspond to the quantitative results summarized in Table~\ref{tab:ars_quantitative}, comparing routes generated by Gurobi, CPLEX, and D-Wave CQM under identical environmental and constraint settings. Insets (a) and (b) highlight local regions where pronounced zigzag patterns appear. \newrevise{Inset (c) presents the raw CMEMS sea-ice-thickness field without the 
hexagonal lattice overlay, improving visibility of the underlying spatial structure 
used during optimization.}}
\label{fig:ars_routes}
\end{figure*}

Each route was analyzed using three physically interpretable indicators:
\begin{itemize}
    \item Total Route Length (km): calculated as the cumulative great-circle distance between consecutive H3 centroids using Equation~\eqref{eq:geodesic_route}.
    \item Zigzag Proxy (\%): quantifies directional smoothness as 
    $\zeta=\sum_{(i,j,k)}\!\big(1-\cos\theta_{ijk}\big)$,  
    where $\theta_{ijk}$ is the turning angle at waypoint $j$; smaller values indicate smoother, more fuel-efficient paths.
    \item CO\textsubscript{2} Estimate (kg): derived from route length and vessel mass as  
    $\mathrm{CO_2}=L_{\mathrm{km}}M_{\mathrm{cargo}}\eta/1,000$,  
    with $M_{\mathrm{cargo}}\!=\!50{,}000$t representing a medium bulk carrier and $\eta\!=\!10$ gCO\textsubscript{2}/t·km following~\cite{ETW2025_ISO14083,JIANG2024177}.  
    This provides a comparative environmental footprint rather than an exact emission prediction.
\end{itemize}
\newrevise{Interpreting zigzag is essential because smoother paths correspond to safer, more fuel-efficient Arctic trajectories, whereas high zigzag values indicate unstable, impractical routes that no vessel could safely follow under ice conditions.}
The numerical outcomes are summarized in Table~\ref{tab:ars_quantitative}, while Fig.~\ref{fig:ars_routes} visualizes the corresponding optimal trajectories.  
Across all problem sizes, D-Wave CQM consistently produces the lowest objective values, shortest total distances, and smoothest curvature.  
For the largest 7,884-node graph, CQM reduces total length by~1.3\% and the zigzag proxy by~9-10\% compared with CPLEX, translating to a~27,000kg decrease in estimated CO\textsubscript{2}.  
Both Gurobi and CPLEX reach comparable optima only after extended runtime, often approaching their 4,200s limit, while the CQM achieves near-identical energies within~30s.

\begin{table*}[!t]
\caption{Quantitative comparison of solver performance for Arctic route optimization across increasing graph scales (small, medium, and large). 
Metrics include objective value, selected nodes, total distance, zigzag proxy, estimated CO\textsubscript{2} emissions, and solver runtime. 
The D-Wave CQM achieves lower objective values and smoother routes within seconds, whereas Gurobi and CPLEX require longer runtimes to achieve comparable quality. 
(\textbf{↓}) indicates that lower values are better.}
\label{tab:ars_quantitative}
\centering
\begin{tabular*}{\tblwidth}{@{} LLLLLLLLL@{} }
\toprule
\textbf{Solver} & \textbf{Nodes} & \textbf{\makecell{Number of\\Quadratic Terms}} &
\textbf{\makecell{Objective\\Value (↓)}} & \textbf{\makecell{Selected\\Nodes (↓)}} &
\textbf{Km (↓)} & \textbf{\makecell{zigzag\\Proxy (\%) (↓)}} & \textbf{CO\textsubscript{2} (kg) (↓)} & \textbf{Time (s) (↓)} \\
\midrule
Gurobi & \multirow{3}{*}{2,045} & \multirow{3}{*}{5,596} & 1,618.8303 & 84 & 1,483.14 & 25.77 & 741{,}569 & 4,200 \\
CPLEX  &  &  & 1,613.0589 & 84 & 1,483.03 & 25.78 & 741{,}514 & 4,200 \\
\textbf{CQM} &  &  & \textbf{1,608.4692} & 82 & \textbf{1,448.62} & \textbf{21.81} & \textbf{724{,}310} & \textbf{5} \\
\midrule
Gurobi & \multirow{3}{*}{5,130} & \multirow{3}{*}{14,234} & 2,583.7516 & 174 & 3,165.44 & 34.33 & 1{,}582{,}719 & 4,200 \\
CPLEX  &  &  & 2,572.5410 & 173 & 3,148.78 & 32.40 & 1{,}573{,}889 & 4,200 \\
\textbf{CQM} &  &  & \textbf{2,566.8319} & 172 & \textbf{3,130.94} & \textbf{29.84} & \textbf{1{,}565{,}471} & \textbf{15} \\
\midrule
Gurobi & \multirow{3}{*}{7,884} & \multirow{3}{*}{21,959} & 5,119.9791 & 225 & 4,097.32 & 35.88 & 2{,}048{,}659 & 4,200 \\
CPLEX  &  &  & 5,102.3211 & 224 & 4,079.49 & 34.90 & 2{,}039{,}746 & 4,200 \\
\textbf{CQM} &  &  & \textbf{5,028.4132} & 222 & \textbf{4,043.29} & \textbf{32.28} & \textbf{2{,}021{,}646} & \textbf{30} \\
\bottomrule
\end{tabular*}
\end{table*}

Several noteworthy patterns emerge:
\begin{enumerate}
\item Trajectory smoothness: Classical solvers, when stopped early, satisfy flow constraints but generate piecewise-linear segments with sharp angular deviations.  
CQM’s quadratic energy minimization naturally penalizes significant directional changes, producing smoother trajectories that align with physically plausible vessel motion.
\item Environmental adaptation: The hybrid solver tends to navigate corridors of thinner ice and lower drift magnitude.  
Visualization in Fig.~\ref{fig:ars_routes} shows CQM paths following low-risk channels within the CMEMS field, reflecting adaptive environmental awareness embedded in the cost function.
\item Runtime scaling: Even as graph resolution triples, CQM runtime grows sub-linearly, confirming robust hybrid orchestration.  
This allows fine-scale Arctic planning without exponential computational overhead.
\end{enumerate}

\subsection{Analysis for Budget Recommendation}
\label{subsec:exp4}

To investigate whether extending the solver’s runtime beyond its internal hybrid allocation (5-30s) improves convergence quality.

\begin{table*}
\caption{CQM solver performance across extended runtime limits. The results indicate that solution quality stabilizes at the solver’s auto time limits (5s for small, 15s for medium, and 30s for large graphs). Extending the runtime to 60 to 150s yields no meaningful improvement, only minor random fluctuations, demonstrating that the default auto time limit offers the most efficient and effective configuration. (\textbf{↓}) indicates that lower values are better.}
\label{tab:cqm_budget}
\centering
\begin{tabular*}{\tblwidth}{@{} LLLLLLL@{} }
\toprule
\textbf{Solver} & \textbf{Objective Value (↓)} & \textbf{Selected Nodes (↓)} & 
\textbf{Km (↓)} & \textbf{zigzag Proxy (\%) (↓)} & \textbf{CO\textsubscript{2} (kg) (↓)} & \textbf{Time (s)} \\
\midrule
\multirow{5}{*}{CQM}
 & 5,028.4132 & 222 & 4,043.2914 & 32.2754 & 2{,}021{,}645.71 & 30 \\
 & 5,028.4132 & 222 & 4,043.2914 & 32.2754 & 2{,}021{,}645.71 & 60 \\
 & 5,030.4076 & 222 & 4,043.3109 & 32.7742 & 2{,}021{,}655.46 & 90 \\
 & 5,031.7375 & 222 & 4,043.3238 & 32.7852 & 2{,}021{,}661.91 & 120 \\
 & 5,028.4132 & 222 & 4,043.2914 & 32.2754 & 2{,}021{,}645.71 & 150 \\
\bottomrule
\end{tabular*}
\end{table*}

\begin{figure*}[!t]
\centering
\includegraphics[width=\textwidth]{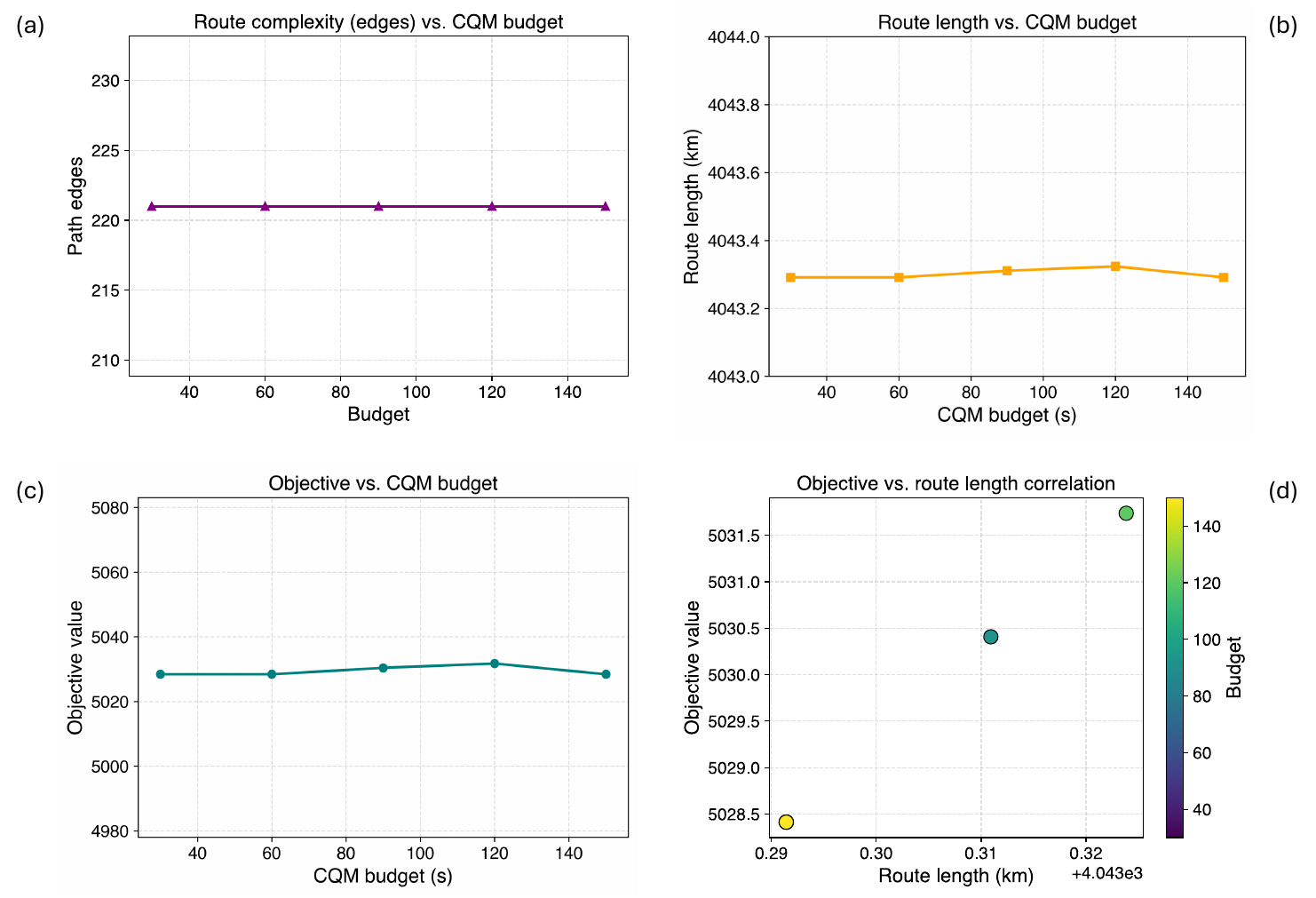}
\caption{Comparative analysis of the D-Wave CQM solver under increasing runtime budgets (30-150s). Subfigures (a)-(d) demonstrate that solver performance converges structurally and numerically within the default 30s hybrid allocation. (a)~Route complexity remains constant at approximately 221 path edges, corresponding to 222 selected nodes (as shown in Table~\ref{tab:cqm_budget}), confirming identical topological reconstruction across all budgets. (b)~Path length and derived navigational metrics (zigzag proxy and CO$_2$ emissions) remain effectively unchanged, indicating geometric and energetic stability. (c)~Objective trajectories exhibit only marginal improvement ($\approx$0.16\%) despite a fivefold increase in runtime. (d)~Objective and length correlation shows a fully flattened gradient beyond 30s, implying solver saturation and diminishing returns. The data points corresponding to the 30s and 150s budgets nearly coincide, confirming that extended runtimes yield no further geometric or energetic improvement.}
\label{fig:budget_analysis}
\end{figure*}

Table~\ref{tab:cqm_budget} confirms that all performance indicators, objective value, selected nodes, path length, zigzag proxy, and CO\textsubscript{2} estimate, remain effectively constant across the 30-150s range.  
Objective values fluctuate within less than~0.05\%, and the number of selected nodes remains exactly~222 for all runs, demonstrating complete structural consistency of the optimized route.  
These quantitative findings correspond directly to the visual trends in Fig.~\ref{fig:budget_analysis}:

\begin{itemize}
\item Panel (a), route complexity:
The number of active path edges stabilizes at approximately~221 across all runtime budgets, confirming that the solver converges to a single connected solution topology by 30s.  
No additional nodes or branches appear at longer durations, indicating that topological exploration is already complete during the default hybrid cycle.

\item Panel (b), geometric and energetic metrics:
The total route length and its derived metrics, zigzag proxy and CO\textsubscript{2} emissions, remain virtually unchanged.  
The small horizontal overlap of all curves highlights geometric invariance, with deviations below~0.01\%.  
This stability implies that once the hybrid solver identifies the low-energy feasible basin, subsequent samples merely reproduce the same route geometry.

\item Panel (c), objective trajectory:  
The total objective exhibits only marginal stochastic variation ($\sim$0.16\%) despite a fivefold increase in runtime.  
Such small fluctuations arise from minor differences in floating-point rounding and random seed initialization within the hybrid orchestration, rather than from genuine improvements in solution quality.

\item Panel (d), objective-length correlation:
The correlation curve between objective value and total distance completely flattens beyond 30s, signaling solver saturation.  
The near-perfect overlap between the 30s and 150s data points confirms that additional sampling yields no further geometric or energetic gain.
\end{itemize}
  
Together, these panels verify that the D-Wave hybrid CQM solver reaches full convergence by its internal default allocation (approximately~30s for large graphs).  
The absence of topological change, geometric deviation, or energy reduction across longer runs demonstrates that the hybrid orchestration efficiently balances classical preprocessing and quantum sampling without requiring manual extension.  
Beyond the default time limit, further computation introduces only stochastic jitter rather than new information.

Based on these results, runtime extension beyond the solver’s adaptive limit (5-30s, depending on graph scale) provides no measurable benefit while linearly increasing computational cost.  
For operational Arctic routing deployments, where real-time responsiveness and resource efficiency are critical, the built-in adaptive limit should therefore be retained as the standard configuration.  
This ensures an optimal balance between runtime, energy cost, and solution fidelity while preserving route stability and physical interpretability.

\subsection{Overall Discussion and Key Insights}
\label{subsec:discussion}

Across all experiments, the hybrid D-Wave~CQM solver consistently demonstrated superior time-to-solution and route realism compared to classical solvers.  
Several overarching insights emerge:

\begin{itemize}
\item Computational advantage:  
CQM achieves near-optimal energies within seconds, even for thousands of binary variables, while classical solvers require hours or fail to converge.  
This establishes a practical, not merely theoretical, advantage in turnaround time.

\item Scalability:
Unlike the exponential scaling of branch-and-bound methods, CQM runtime remains nearly constant as quadratic coupling grows.  
This property is crucial for high-resolution Arctic modeling, where each increase in the H3 level doubles or triples the number of variables.

\item Physical interpretability:
The resulting Arctic routes exhibit smoother curvature, continuous flow, and reduced zigzag proxy, key indicators of realistic ship trajectories.  
Such improvements reflect the solver’s inherent capacity to balance nonconvex penalties without producing fragmented or over-penalized paths.

\item Operational relevance:
The budget analysis confirms that hybrid solvers can operate within predictable time windows, making them compatible with real-time or on-demand maritime decision support systems.
\end{itemize}

In summary, the \newrevise{our} framework bridges environmental realism and computational scalability.  
By delivering physically interpretable, low-energy routes within seconds, it offers a viable foundation for next-generation Arctic navigation platforms.  
These experimental findings substantiate the broader claim that hybrid quantum annealing optimization can achieve practical quantum efficiency in high-dimensional, multi-criteria maritime routing problems.

\section{Conclusion and Future Work}
\label{sec:conclusion}
This study presented a comprehensive evaluation of a hybrid quantum annealing optimization framework for Arctic route planning under realistic sea-ice and oceanographic conditions.  
Through sequential experiments ranging from synthetic formulations to real data-driven Arctic scenarios, the proposed model demonstrated that the D-Wave CQM solver consistently achieves near-optimal results within seconds.  
Results show that the CQM consistently attains near-optimal routes within seconds, whereas classical solvers such as Gurobi and CPLEX require longer runtimes to reach comparable or higher-cost solutions.

The optimized routes are physically interpretable, smooth, and energy efficient, with reduced zigzag deviation and lower estimated CO\textsubscript{2} emissions. 
These outcomes confirm that the hybrid quantum approach can balance computational scalability and environmental realism, establishing a practical foundation for real-time Arctic navigation and sustainability management.

Several limitations remain.
The present analysis uses a single temporal snapshot of sea-ice data at H3 level~5, limiting representation of seasonal variability and sub-grid deformation. 
Enhancing the H3 spatial resolution from level 5 to higher levels (e.g., 7 \revise{or} 8) will improve route granularity and coastal detail, although it will proportionally increase the optimization dimensionality and computational demand.
Regulatory factors from the IMO Polar Code and detailed propulsion-based emission models were not yet incorporated, and full quantum advantage is still constrained by hardware connectivity and noise.

Future extensions will embed POLARIS-compliant safety layers~\cite{WANG2025120294}, higher-resolution sea-ice forecasts~\cite{OpenMeteo2025}, and vessel-specific performance models to create an adaptive, time-responsive routing system. 
The ongoing D-Wave Advantage2\texttrademark~architecture, exceeding 4{,}400~qubits with enhanced connectivity, provides a promising pathway toward scaling~\cite{DWaveAdvantage2_2025} these hybrid formulations to tens of thousands of binary variables for next-generation Arctic navigation.
According to D-Wave’s published roadmap, the forthcoming generation of annealing processors is projected to exceed 7,000 qubits with higher order (20-way) connectivity,
providing a viable path toward tackling tens of thousands of binary variables within hybrid quantum-classical workflows~\cite{DWaveRoadmap_2025}.

Overall, this research demonstrates that hybrid quantum annealing constrained optimization can deliver reliable, near-optimal Arctic routes efficiently, bridging scientific modeling, safer route planning, and sustainable polar transportation.

% \appendix
% \section{My Appendix}
% Appendix sections are coded under \verb+\appendix+.

\printcredits

\bibliographystyle{elsarticle-num}
% \bibliographystyle{cas-model2-names}

\bibliography{cas-refs}

@article{Zhang2025_NatureCO2,
  title        = {Arctic Sea Route access reshapes global shipping carbon emissions},
  author       = {Zhao, Peng and Li, Yifan and Zhang, Chao and others},
  journal      = {Nature Communications},
  volume       = {16},
  pages        = {8431},
  year         = {2025},
  publisher    = {Nature Publishing Group},
  doi          = {10.1038/s41467-025-64437-4},
}

@article{Kozera2025_Grassroots,
  title        = {The Role and Importance of the Arctic and its Sea Route in International Economic Relations},
  author       = {Kozera, {\L}ukasz and K{\l}aczy\'{n}ski, Rafa\l{}},
  journal      = {Grassroots Journal of Natural Resources},
  volume       = {8},
  number       = {1},
  pages        = {709--736},
  year         = {2025},
  doi          = {10.33002/nr2581.6853.080130},
}

@article{Rodriguez2025_AIS,
    author = {Jorge P. Rodríguez and Konstantin Klemm and Carlos M. Duarte and Víctor M. Eguíluz},
    title = {Shipping traffic through the Arctic Ocean: Spatial distribution, seasonal variation, and its dependence on the sea ice extent},
    journal = {iScience},
    volume = {27},
    number = {7},
    pages = {110236},
    year = {2024},
    issn = {2589-0042},
    doi = {10.1016/j.isci.2024.110236},
}

@article{Zhang2024_OceanEng,
    title = {Influence of sea ice on ship routes and speed along the Arctic Northeast Passage},
    journal = {Ocean \& Coastal Management},
    volume = {256},
    pages = {107320},
    year = {2024},
    issn = {0964-5691},
    doi = {10.1016/j.ocecoaman.2024.107320},
    author = {Yaqing Shu and Hailong Cui and Lan Song and Langxiong Gan and Sheng Xu and Jie Wu and Chunmiao Zheng},
}

@article{Kotovirta2023_Review,
    title = {Pathfinding and optimization for vessels in ice: A literature review},
    journal = {Cold Regions Science and Technology},
    volume = {211},
    pages = {103876},
    year = {2023},
    issn = {0165-232X},
    doi = {10.1016/j.coldregions.2023.103876},
    author = {Trung Tien Tran and Thomas Browne and Mashrura Musharraf and Brian Veitch},
}

@ARTICLE{Quirante2024_MIQP,
  author={Quirynen, Rien and Safaoui, Sleiman and Di Cairano, Stefano},
  journal={IEEE Transactions on Control Systems Technology}, 
  title={Real-Time Mixed-Integer Quadratic Programming for Vehicle Decision-Making and Motion Planning}, 
  year={2025},
  volume={33},
  number={1},
  pages={77-91},
  doi={10.1109/TCST.2024.3449703}}

@Article{AlKhayyal2022_Piecewise,
    AUTHOR = {Alkhalifa, Loay and Mittelmann, Hans},
    TITLE = {New Algorithm to Solve Mixed Integer Quadratically Constrained Quadratic Programming Problems Using Piecewise Linear Approximation},
    JOURNAL = {Mathematics},
    VOLUME = {10},
    YEAR = {2022},
    NUMBER = {2},
    ARTICLE-NUMBER = {198},
    ISSN = {2227-7390},
    DOI = {10.3390/math10020198}
}

@article{Huang2020_ShipIceCFD,
    title = {CFD-DEM based full-scale ship-ice interaction research under FSICR ice condition in restricted brash ice channel},
    journal = {Cold Regions Science and Technology},
    volume = {194},
    pages = {103454},
    year = {2022},
    issn = {0165-232X},
    doi = {10.1016/j.coldregions.2021.103454},
    author = {Jianing Zhang and Yi Zhang and Yuchen Shang and Qiang Jin and Lei Zhang}
}

@misc{DWave2025_CQMDocs,
  author={{D-Wave Systems}},
  title={Ocean Documentation: Constrained Quadratic Models},
  year={2025},
  url={https://docs.dwavequantum.com}
}

@ARTICLE{Pelofske2025_Hybrid,
  author={Osaba, Eneko and Miranda-Rodriguez, Pablo},
  journal={IEEE Access}, 
  title={D-Wave’s Nonlinear-Program Hybrid Solver: Description and Performance Analysis}, 
  year={2025},
  volume={13},
  number={},
  pages={4724-4736},
  doi={10.1109/ACCESS.2025.3525620}
}

@article{Quinton2025_QAReview,
  title        = {Quantum annealing applications, challenges and limitations for optimisation problems compared to classical solvers},
  author       = {Quinton, F. A. and Myhr, P. A. S. and Barani, M. and others},
  journal      = {Scientific Reports},
  volume       = {15},
  pages        = {12733},
  year         = {2025},
  doi          = {10.1038/s41598-025-96220-2},
}

@article{Szal2025_TransNav,
  title        = {Benchmarking the Maritime Inventory Routing Problem on a Quantum Annealing-Hybrid System},
  author       = {Szal, O. and Rubbert, S. and Rizvanolli, A.},
  journal      = {TransNav: International Journal on Marine Navigation and Safety of Sea Transportation},
  volume       = {13},
  number       = {1},
  pages        = {97--102},
  year         = {2019},
  doi          = {10.12716/1001.13.01.14},
}

@article{10632778,
  author={Malcolm, John D. and Roth, Alexander and Radic, Mladjan and Martín-Ramiro, Pablo and Oillarburu, Jon and Aizpurua, Borja and Orús, Román and Mugel, Samuel},
  journal={IEEE Transactions on Quantum Engineering}, 
  title={Multidisk Clutch Optimization Using Quantum Annealing}, 
  year={2024},
  volume={5},
  number={},
  pages={1-10},
  doi={10.1109/TQE.2024.3441229}}

@article{Theocharis2023_Frontiers,
  title        = {Arctic weather routing: a review of ship performance models and ice routing algorithms},
  author       = {Liu, Qiang and Wang, Yue and Zhang, Rui and Yan, Hao and Xu, Jie and Guo, Yong},
  journal      = {Frontiers in Marine Science},
  volume       = {10},
  pages        = {1190164},
  year         = {2023},
  doi          = {10.3389/fmars.2023.1190164},
  publisher    = {Frontiers in Marine Science}
}

@article{Saebi2025_Ballast,
  title        = {Network analysis of ballast-mediated species transfer reveals important introduction and dispersal patterns in the Arctic},
  author       = {Saebi, Mehdi and Xu, Jing and Curasi, Steven R. and others},
  journal      = {Scientific Reports},
  volume       = {10},
  pages        = {19558},
  year         = {2020},
  doi          = {10.1038/s41598-020-76602-4},
  publisher    = {Nature Publishing Group}
}

@article{Poo2024_TradeType,
    author = {Mark Ching-Pong Poo and Zaili Yang and Yui-yip Lau and Pisit Jarumaneeroj},
    title = {Assessing the impact of Arctic shipping routes on the global container shipping network’s connectivity},
    journal = {Polar Geography},
    volume = {47},
    number = {3},
    pages = {219--239},
    year = {2024},
    publisher = {Taylor \& Francis},
    doi = {10.1080/1088937X.2024.2399775},
}

@ARTICLE{Quirynen2024_ControlMIQP,
  author={Reiter, Rudolf and Quirynen, Rien and Diehl, Moritz and Di Cairano, Stefano},
  journal={IEEE Transactions on Control Systems Technology}, 
  title={Equivariant Deep Learning of Mixed-Integer Optimal Control Solutions for Vehicle Decision Making and Motion Planning}, 
  year={2025},
  volume={33},
  number={4},
  pages={1270-1284},
  doi={10.1109/TCST.2024.3400571}
}

@article{Aksenov2023_Risks,
  title        = {New insights into projected Arctic sea road: operational risks, economic values, and policy implications},
  author = {Li, Xueke and Lynch, Amanda H.},
  journal      = {Climatic Change},
  volume       = {176},
  number       = {4},
  pages        = {30},
  year         = {2023},
  doi          = {10.1007/s10584-023-03505-4},
  publisher    = {Springer Nature}
}

@inproceedings{10.1145/3760622.3760626,
    author = {Chow, Elle Wing Ho and Ho, George To Sum and Tang, Valerie and Tam, Manviel Man Fei},
    title = {Utilizing Quantum Annealing to Address Vehicle Routing Challenges in Cold Chain Logistics},
    year = {2025},
    isbn = {9798400710452},
    publisher = {Association for Computing Machinery},
    address = {New York, NY, USA},
    doi = {10.1145/3760622.3760626},
    booktitle = {Proceedings of the 2025 9th International Conference on Intelligent Systems, Metaheuristics \& Swarm Intelligence},
    pages = {137–142},
    numpages = {6},
}

@article{Wang2024_YearRound,
  author       = {Zhao, P. and Li, Y. and Zhang, Y.},
  title        = {Ships are projected to navigate whole year-round along the North Sea route by 2100},
  journal      = {Communications Earth \& Environment},
  volume       = {5},
  pages        = {407},
  year         = {2024},
  publisher    = {Nature Portfolio},
  doi          = {10.1038/s43247-024-01557-7},
}

@article{Wang2025_ShippingPolicy,
    title = {The strategic insights of Arctic sea routes for the sustainable development of Taiwan's shipping industry},
    journal = {Transport Policy},
    volume = {159},
    pages = {190-200},
    year = {2024},
    issn = {0967-070X},
    author = {Yung-Sheng Chen and Po-Hung Chen and Chun-Hao Jung and Tsai-Ling Chang and Jia-An Ye and Ta-Kang Liu},
    doi = {10.1016/j.tranpol.2024.10.021},
}

@misc{Uber2018H3,
  title        = {H3: Uber’s Hexagonal Hierarchical Spatial Index},
  author       = {{Uber Technologies, Inc.}},
  year         = {2018},
  url          = {https://h3geo.org/docs/core-library/restable/},
  note         = {Accessed: 2025-10-20}
}

@misc{CMEMS_MDS_00016,
  title        = {Global Ocean Physics Analysis and Forecast},
  author       = {{E.U. Copernicus Marine Service Information (CMEMS)}},
  year         = {2025},
  publisher    = {Marine Data Store (MDS)},
  doi          = {10.48670/moi-00016},
  note         = {Accessed: 20-Oct-2025},
}

@misc{GSHHS_dataset,
  title        = {Global Self-consistent, Hierarchical, High-resolution Geography Database (GSHHG)},
  author       = {Wessel, Paul and Smith, Walter H. F.},
  year         = {2023},
  url          = {https://www.soest.hawaii.edu/pwessel/gshhg/},
  note         = {Version 2.3.7, Accessed: 2025-10-20}
}

@article{WANG2025120294,
    title = {Incremental route planning based on daily risk assessment for Arctic navigation},
    journal = {Ocean Engineering},
    volume = {320},
    pages = {120294},
    author = {Bo Wang and Yuanmin Liu and Wen Dai and Jixuan Li},
    year = {2025},
    issn = {0029-8018},
    doi = {10.1016/j.oceaneng.2025.120294},
}

@misc{OpenMeteo2025,
  author       = {{Open-Meteo}},
  title        = {Open-Meteo Weather API: Free Weather Forecast API for Non-Commercial Use},
  howpublished = {\url{https://open-meteo.com/en/docs#location_and_time}},
  note         = {Accessed: 28-Oct-2025},
  year         = {2025}
}

@misc{DWaveAdvantage2_2025,
  author       = {{D\textendash Wave Quantum Inc.}},
  title        = {D\textendash Wave Announces General Availability of Advantage2\texttrademark\ Quantum Computer — Production-Ready 4{,}400{+} Qubit Annealing Quantum Computer},
  howpublished = {Press Release, May-20-2025, Palo~Alto,~CA}, 
  note         = {Accessed: 28-Oct-2025},
  year         = {2025}
}

@techreport{DWave_HybridSolvers_2022,
  author       = {{D-Wave Systems Inc.}},
  title        = {Hybrid Solvers for Quadratic Optimization},
  year         = {2022},
  institution  = {D-Wave Systems Inc.},
  type         = {White Paper},
  url          = {https://www.dwavesys.com/media/soxph512/hybrid-solvers-for-quadratic-optimization.pdf},
  note         = {Accessed: 2025-11-04}
}

@misc{DWaveRoadmap_2025,
  author       = {{D\textendash Wave Quantum Inc.}},
  title        = {Clarity Roadmap: Advancing Quantum Annealing and Hybrid Quantum Computing Systems},
  howpublished = {Technical Roadmap, D\textendash Wave Quantum Inc., 2025},
  url          = {https://www.dwavequantum.com/media/xvjpraig/clarity-roadmap_digital_v2.pdf},
  note         = {Accessed: 28-Oct-2025},
  year         = {2025}
}

@article{Krauss2020,
  author={Krauss, Thomas and McCollum, Joey},
  journal={IEEE Transactions on Quantum Engineering}, 
  title={Solving the Network Shortest Path Problem on a Quantum Annealer}, 
  year={2020},
  volume={1},
  number={},
  pages={1-12},
  doi={10.1109/TQE.2020.3021921}
}

@article{Vaidheeswaran2025HexGridRL,
  author       = {V. Vaidheeswaran and others},
  title        = {Goal-Conditioned Reinforcement Learning for Data-Driven Maritime Navigation using Hexagonal Grids},
  journal      = {arXiv preprint arXiv:2509.01838},
  year         = {2025},
  url          = {https://arxiv.org/abs/2509.01838}
}

@article{woodruff2019dynamic,
  author = {Woodruff, David L. and Deride, Julio and Smith, Jean-Paul Watson},
  title = {Dynamic Economic Dispatch Using Complementary Quadratic Programming},
  journal = {Energy},
  year = {2019},
  volume = {166},
  pages = {755--764},
  doi = {10.1016/j.energy.2018.10.119}
}

@article{ROGNE2025121233,
    title = {Regression-based identification of quadratic transfer functions},
    journal = {Ocean Engineering},
    volume = {333},
    pages = {121233},
    year = {2025},
    issn = {0029-8018},
    doi = {10.1016/j.oceaneng.2025.121233},
    author = {Øyvind Ygre Rogne and Thomas Sauder},
}

@article{zhang2023changing,
  author = {Zhang, Yu and Sun, Xiaopeng and Zha, Yufan and Wang, Kun and Chen, Changsheng},
  title = {Changing Arctic Northern Sea Route and Transpolar Sea Route: A Prediction of Route Changes and Navigation Potential before Mid-21st Century},
  journal = {Journal of Marine Science and Engineering},
  year = {2023},
  volume = {11},
  number = {12},
  pages = {2340},
  doi = {10.3390/jmse11122340},
}

@article{li2024feasibility,
    author = {Li, Tongtong and Wang, Yangjun and Li, Yan and Wang, Bin and Liu, Quanhong and Chen, Xi},
    title = {Feasibility of the Northern Sea Route: Impact of Sea Ice Thickness Uncertainty on Navigation},
    journal = {Journal of Marine Science and Engineering},
    year = {2024},
    volume = {12},
    number = {7},
    pages = {1078},
    doi = {10.3390/jmse12071078},
}

@article{song2022routeview,
    author = {Song, Chengming and Zhang, Fang and Chen, Jing and Yang, Chunxia and Yap, John},
    title = {Routeview: an intelligent route planning system for ships sailing through Arctic ice zones based on big Earth data},
    journal = {International Journal of Digital Earth},
    year = {2022},
    volume = {15},
    number = {1},
    pages = {1850--1873},
    doi = {10.1080/17538947.2022.2126016},
}

@article{bousquin2021discrete,
    author = {Bousquin, Justin},
    title = {Discrete Global Grid Systems as Scalable Geospatial Frameworks for Characterizing Coastal Environments},
    journal = {Environmental Modelling \& Software},
    year = {2021},
    volume = {146},
    pages = {105210},
    doi = {10.1016/j.envsoft.2021.105210},
}

@article{spadon2025goal,
    author = {Spadon, Gabriel},
    title = {Goal-Conditioned Reinforcement Learning for Data-Driven Maritime Navigation},
    journal = {arXiv preprint arXiv:2509.01838},
    year = {2025},
    url = {https://arxiv.org/abs/2509.01838}
}

@article{SalvagninRobertiFischetti2024,
  author    = {D. Salvagnin and R. Roberti and M. Fischetti},
  title     = {A fix-propagate-repair heuristic for mixed integer programming},
  journal   = {Mathematical Programming Computation},
  year      = {2024},
  doi       = {10.1007/s12532-024-00269-5}
}

@article{HojnyJoormannLuthenSchmidt2021,
  author    = {C. Hojny and I. Joormann and H. Lüthen and M. Schmidt},
  title     = {Mixed-integer programming techniques for the connected Max-k-Cut problem},
  journal   = {Mathematical Programming Computation},
  year      = {2021},
  doi       = {10.1007/s12532-020-00186-3}
}

@article{NaikBemporad2021,
  author    = {Naik, Vihangkumar V. and Bemporad, Alberto},
  title     = {Exact and Heuristic Methods with Warm-start for Embedded Mixed-Integer Quadratic Programming Based on Accelerated Dual Gradient Projection},
  journal   = {arXiv preprint arXiv:2101.09264},
  year      = {2021},
  url       = {https://arxiv.org/abs/2101.09264}
}

@article{TangXuAmosKolter2024,
  title        = {Learning to Optimize for Mixed-Integer Non-linear Programming with Feasibility Guarantees},
  author       = {Tang, Bo and Khalil, Elias B. and Drgo{\v{n}}a, J{\'a}n},
  journal      = {arXiv preprint arXiv:2410.11061},
  year         = {2025},
  month        = {May},
  url          = {https://arxiv.org/abs/2410.11061},
  note         = {arXiv:2410.11061 [cs.LG]},
  institution  = {University of Toronto and Johns Hopkins University}
}

@techreport{ETW2025_ISO14083, author = {{EcoTransIT World Consortium}}, institution = {EcoTransIT} , title = {EcoTransIT World Methodology Report (Version 4), ISO 14083 aligned}, year = {2025}, url = {https://www.ecotransit.org/wp-content/uploads/EcoTransIT_World_Methodology_Report_Version_4_ISO14083.pdf}
}

@article{choic2023,
    author = {Choi, Gwang-Hyeok and Lee, Wonhee and Kim, Tae-wan},
    title = {Voyage optimization using dynamic programming with initial quadtree based route},
    journal = {Journal of Computational Design and Engineering},
    volume = {10},
    number = {3},
    pages = {1185-1203},
    year = {2023},
    month = {06},
    issn = {2288-5048},
    doi = {10.1093/jcde/qwad055},
}

@article{CHEN2025120956,
title = {Intelligent ship route planning via an A{$^\ast$} search model enhanced double-deep Q-network},
journal = {Ocean Engineering},
volume = {327},
pages = {120956},
year = {2025},
issn = {0029-8018},
doi = {https://doi.org/10.1016/j.oceaneng.2025.120956},
author = {Xinqiang Chen and Ruiyang Hu and Kai Luo and Huafeng Wu and Salvatore Antonio Biancardo and Yiwen Zheng and Jiangfeng Xian},
}

@INPROCEEDINGS{7526551,
  author={Takapoui, Reza and Moehle, Nicholas and Boyd, Stephen and Bemporad, Alberto},
  booktitle={2016 American Control Conference (ACC)}, 
  title={A simple effective heuristic for embedded mixed-integer quadratic programming}, 
  year={2016},
  volume={},
  number={},
  pages={5619-5625},
  doi={10.1109/ACC.2016.7526551}}

@article{lwxq-4myj,
  title = {Slack-variable approach for variational quantum semidefinite programming},
  author = {Chen, Jingxuan and Westerheim, Hanna and Holmes, Zo\"e and Luo, Ivy and Nuradha, Theshani and Patel, Dhrumil and Rethinasamy, Soorya and Wang, Kathie and Wilde, Mark M.},
  journal = {Phys. Rev. A},
  volume = {112},
  issue = {2},
  pages = {022607},
  numpages = {52},
  year = {2025},
  month = {Aug},
  publisher = {American Physical Society},
  doi = {10.1103/lwxq-4myj},
}

@article{s20061550,
    AUTHOR = {Silva Junior, Andouglas Gonçalves da and Santos, Davi Henrique dos and Negreiros, Alvaro Pinto Fernandes de and Silva, João Moreno Vilas Boas de Souza and Gonçalves, Luiz Marcos Garcia},
    TITLE = {High-Level Path Planning for an Autonomous Sailboat Robot Using Q-Learning},
    JOURNAL = {Sensors},
    VOLUME = {20},
    YEAR = {2020},
    NUMBER = {6},
    ARTICLE-NUMBER = {1550},
    PubMedID = {32168774},
    ISSN = {1424-8220},
    DOI = {10.3390/s20061550}
}

@article{Glover2022QUBO,
  author    = {Fred Glover and Gary Kochenberger and Rick Hennig and others},
  title     = {Quantum Bridge Analytics I: A Tutorial on Formulating and Using QUBO Models},
  journal   = {Annals of Operations Research},
  volume    = {314},
  pages     = {141--183},
  year      = {2022},
  doi       = {10.1007/s10479-022-04634-2},
  publisher = {Springer},
}

@article{ROVENSKAYA2024103446,
    title = {Future scenarios of commercial freight shipping in the Euro-Asian Arctic},
    journal = {Futures},
    volume = {163},
    pages = {103446},
    year = {2024},
    issn = {0016-3287},
    doi = {https://doi.org/10.1016/j.futures.2024.103446},
    author = {Elena Rovenskaya and Nikita Strelkovskii and Dmitry Erokhin and Leena Ilmola-Sheppard},
}

@article{Beyaztas2025,
  author    = {Beyaztas, U. and Shang, H. L. and Saricam, S.},
  title     = {Penalized function-on-function linear quantile regression},
  journal   = {Computational Statistics},
  volume    = {40},
  pages     = {301--329},
  year      = {2025},
  doi       = {10.1007/s00180-024-01494-1},
}

@article{XIN2025104660,
    title = {Shape distortion analysis of hexagonal discrete global grid systems based on standard deviation of neighbor distance},
    journal = {International Journal of Applied Earth Observation and Geoinformation},
    volume = {141},
    pages = {104660},
    year = {2025},
    issn = {1569-8432},
    doi = {10.1016/j.jag.2025.104660},
    author = {Zhang Xin and Cao Yibing and Li Tingting},
}

@article{JIANG2024177,
    title = {Quantification of CO2 emissions in transportation: An empirical analysis by modal shift from road to waterway transport in Zhejiang, China},
    journal = {Transport Policy},
    volume = {145},
    pages = {177-186},
    year = {2024},
    issn = {0967-070X},
    doi = {10.1016/j.tranpol.2023.10.026},
    author = {Meizhi Jiang and Benmei Wang and Yingjun Hao and Shijun Chen and Yuanqiao Wen and Zaili Yang},
}

@inproceedings{10.1007/978-3-319-93031-2_43,
    author = {Bonami, Pierre and Lodi, Andrea and Zarpellon, Giulia},
    title = {Learning a Classification of Mixed-Integer Quadratic Programming Problems},
    year = {2018},
    isbn = {978-3-319-93030-5},
    publisher = {Springer-Verlag},
    address = {Berlin, Heidelberg},
    doi = {10.1007/978-3-319-93031-2_43},
    pages = {595–604},
    numpages = {10},
    location = {Delft, The Netherlands},
    booktitle = {Integration of Constraint Programming, Artificial Intelligence, and Operations Research},
}

@article{CHEN2022676,
    title = {Navigability of the Northern Sea Route for Arc7 ice-class vessels during winter and spring sea-ice conditions},
    journal = {Advances in Climate Change Research},
    volume = {13},
    number = {5},
    pages = {676-687},
    year = {2022},
    issn = {1674-9278},
    doi = {10.1016/j.accre.2022.09.005},
    author = {Shi-Yi CHEN and Stefan Kern and Xin-Qing LI and Feng-Ming HUI and Yu-Fang YE and Xiao Cheng},
}

@Article{rs16162983,
    AUTHOR = {Kacimi, Sahra and Kwok, Ron},
    TITLE = {Two Decades of Arctic Sea-Ice Thickness from Satellite Altimeters: Retrieval Approaches and Record of Changes (2003–2023)},
    JOURNAL = {Remote Sensing},
    VOLUME = {16},
    YEAR = {2024},
    NUMBER = {16},
    ARTICLE-NUMBER = {2983},
    ISSN = {2072-4292},
    DOI = {10.3390/rs16162983}
}


% Author biography without author photo.
% Author biography. Author biography. Author biography.

\clearpage
\bio{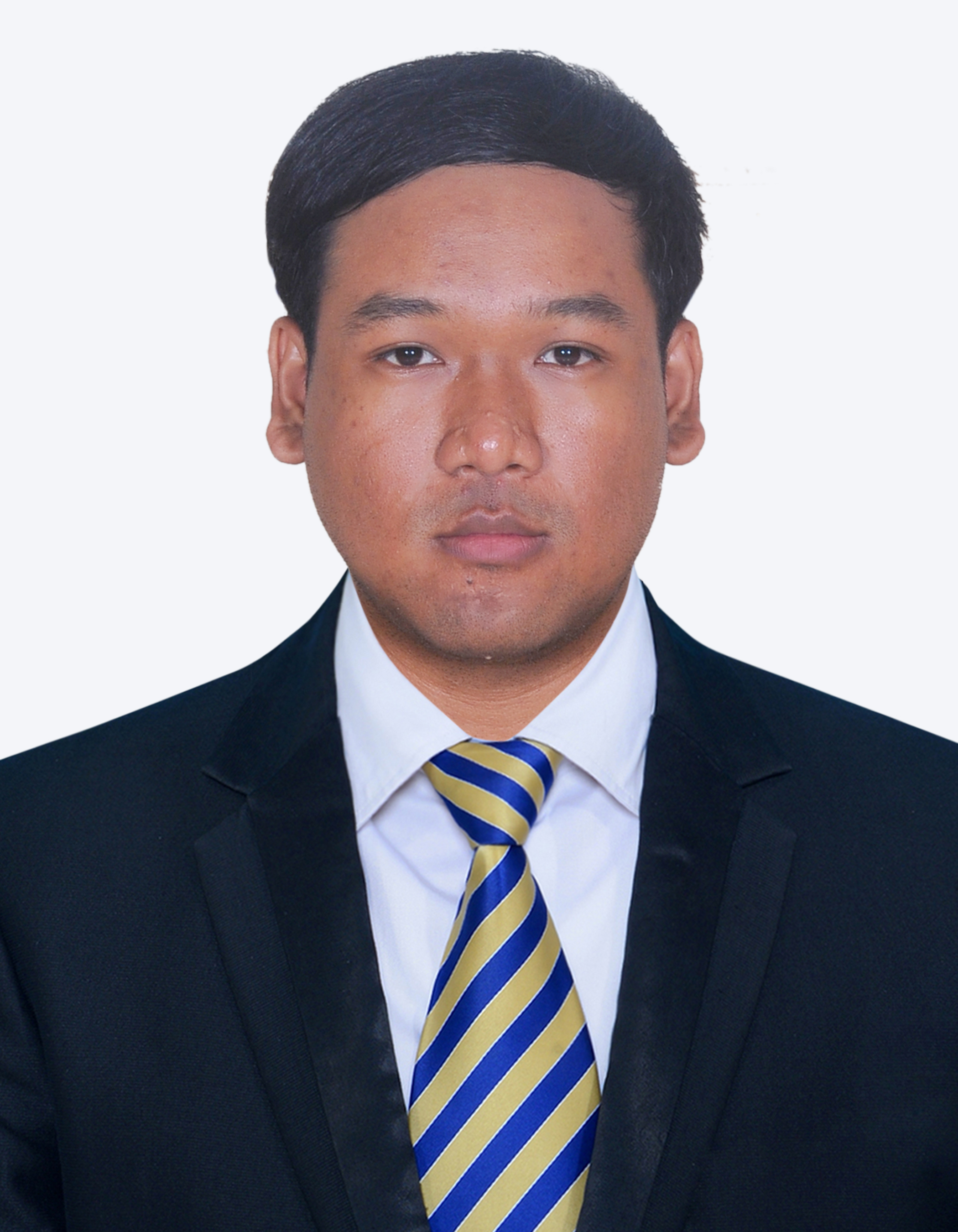}
Tara Kit, February 2020 : Royal University of Phnom Penh, Department of Computer Science.
March 2024 to the present: Pukyong National University. General Graduate School of Artificial Intelligence Convergence. The master's course.
Research interest: Quantum machine learning.
\endbio

\bio{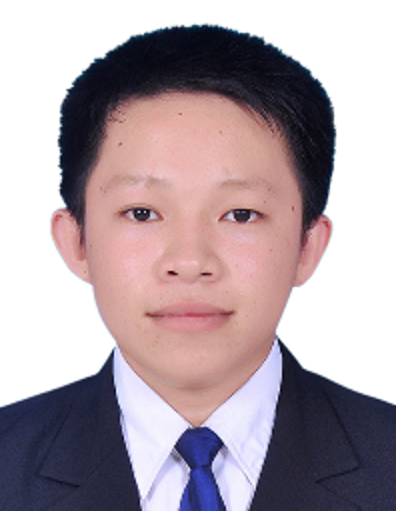}
Kimsay Pov, February 2020 : Royal University of Phnom Penh, Department of Computer Science.
September 2023 to the present: Pukyong National University. General Graduate School of Artificial Intelligence Convergence. The Ph.D. course.
Research interest: Quantum machine learning, deep learning software framework.
\endbio

\bio{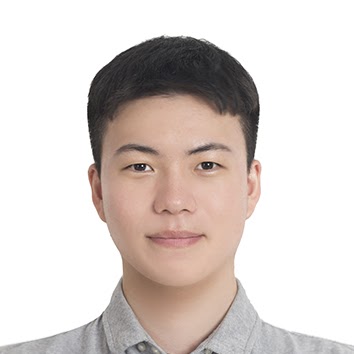}
Myeongseong Go, Feburary 2023 : Pukyong National University of South Korea, Department of Computer Engineering.
March 2023 to present : Pukyong National University. Gerneral Graduate Shcool of Artificial Intelligence Convergence. The Ph.D. course.
Research interest: Quantum machine learning.
\endbio

\bio{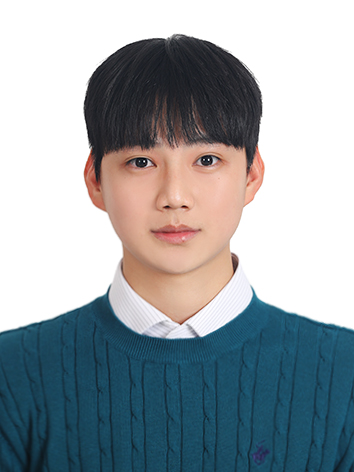}
Leanghok Hour, 2019: Received a Bachelor's degree in Management Information Systems from SETEC Institute, Phnom Penh, Cambodia. He is currently working toward an integrated MS and PhD degree in the Department of AI Convergence at Pukyong National University, Busan, South Korea. Research interests: hybrid quantum-classical compilers, quantum machine learning, quantum software stack.
\endbio

% \newpage
\bio{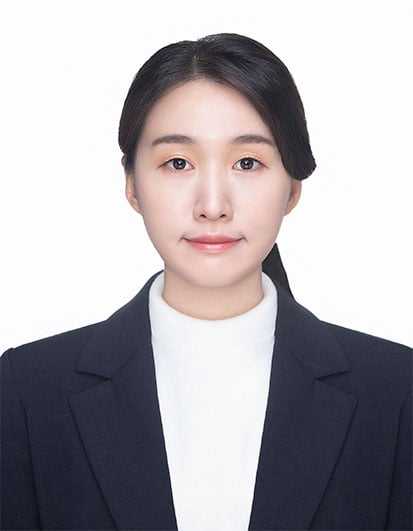}
Arim Ryou, February 2024 : Chungbuk National University, Department of Physics.
March 2024 to the present: Chungbuk National University, Department of Physics. Integrated Master’s and Ph.D. Program.
Research interest: Quantum machine learning, Quantum computing.
\endbio

\bio{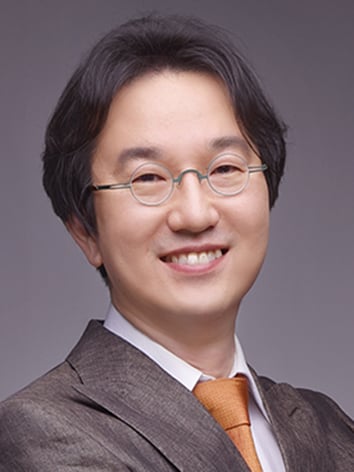}
Kiwoong Kim, February 1995, 1997, 2002: Korea Advanced Institute of Science and Technology (KAIST), Department of Physics, B.S., M.S. and Ph.D. in Solid-state Physics, Respectively. 
April 2002 to Aug. 2020: Research Center Head / Distinguished / Principal Research Scientist, Korea Research Institute of Standards and Science (KRISS). 
June 2006 to Sep. 2007: Research Associate, Department of Physics, Princeton University, U.S.A. 
Sep. 2020 to the present: Professor, Department of Physics, Chungbuk National University, Korea, 2024 to the present: Center Head, Chungbuk Quantum Research Center. Research interest: Quantum sensing, quantum computing, medical imaging, and precision metrology.
\endbio

\newpage
\bio{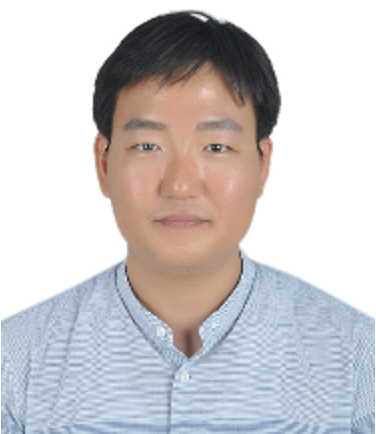}
Tae-Kyung Kim, February 2005: Chungbuk National University Information Mountain, Department of Business Engineering (Doctor of Engineering).
March 2010 to April 2013: Post-doctoral Research Institute of the Korea Research Institute of Biotechnology.
May 2013 to August 2019: Director of SW HRD Center in Korea.
March 2020 to February 2024: An assistant professor at Incheon Talent University.
March 2024 ~ Present: Assistant Professor, Department of Management Information, Chungbuk National University.
Research interest: Artificial Intelligence, Big Data, Quantum Machine Learning.
\endbio

\bio{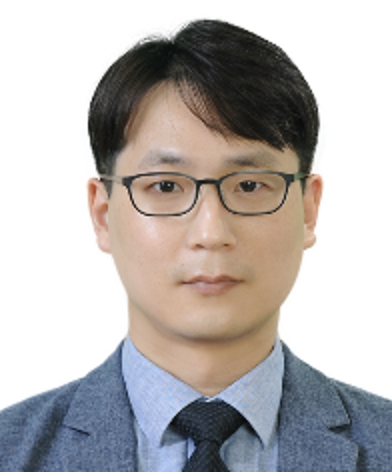}
Youngsun Han, February 2003: Electrical exhibition of Korea University, Department of Electromagnetic Engineering (Engineer).
February 2009: Korea University General University, Department of Electronic Computer Engineering (Doctor of Engineering) at Academy.
June 2009 to February 2011: Samsung LSI Business Department Lead Researcher.
March 2011 to August 2019 : Department of Electronic Engineering, Kyungil University
an adjunct professor.
September 2019 to the present: Computer artificial intelligence engineering at Pukyong National University an undergraduate professor.
Research interest: Quantum computing, compiler design, quantum machine learning.
\endbio

\end{document}